\documentclass[useAMS,usenatbib]{mn2e}

\input{epsf}
\usepackage{graphicx}
\usepackage{deluxetable}

\title[Updating quasar bolometric luminosity corrections]{Updating quasar bolometric luminosity corrections}
\author[J. C. Runnoe et al.]{Jessie C. Runnoe$^{1}$\thanks{E-mail:
jrunnoe@uwyo.edu} , Michael S. Brotherton$^{1}$, and Zhaohui Shang$^{2}$\\
$^{1}$Department of Physics and Astronomy, University of Wyoming, Laramie, WY 82071, USA\\
$^{2}$Department of Physics, Tianjin Normal University, Tianjin 300074, China}
\begin{document}

\date{Preprint 2011 December 6}
\date{Accepted 2012 January 24. Received 2012 January 23; in original form 2011 August 07}

\pagerange{\pageref{firstpage}--\pageref{lastpage}} \pubyear{2011}

\maketitle

\label{firstpage}

\begin{abstract}
Bolometric corrections are used in quasar studies to quantify total energy output based on a measurement of a monochromatic luminosity.  First, we enumerate and discuss the practical difficulties of determining such corrections, then we present bolometric luminosities between 1 $\mu$m and 8 keV rest frame and corrections derived from the detailed spectral energy distributions of 63 bright quasars of low to moderate redshift ($z = 0.03-1.4$).  Exploring several mathematical fittings, we provide practical bolometric corrections of the forms $L_{iso}=\zeta\, \lambda L_{\lambda}$ and log$(L_{iso})=A+B$ log$(\lambda L_{\lambda})$ for $\lambda=$ 1450, 3000, and 5100 \AA, where $L_{iso}$ is the bolometric luminosity calculated under the assumption of isotropy.  The significant scatter in the 5100 \AA\ bolometric correction can be reduced by adding a first order correction using the optical slope, $\alpha_{\lambda,opt}$.  We recommend an adjustment to the bolometric correction to account for viewing angle and the anisotropic emission expected from accretion discs.  For optical/UV monochromatic luminosities, radio-loud and radio-quiet bolometric corrections are consistent within 95\% confidence intervals so we do not make separate radio-loud and radio-quiet corrections.  In addition, we provide several bolometric corrections to the 2-10 keV X-ray luminosity, which are shown to have very large scatter.  Separate radio-loud and radio-quiet corrections are warranted by the X-ray data.
\end{abstract}

\begin{keywords}
galaxies: active Ð quasars: general Ð accretion, accretion discs Ð black hole physics.
\end{keywords}

%INTRODUCTION
%%%%%%%%%%%%%%%%%%%%%%%%%%%%%%%%%%%%%%%%%%%%%%%%%%%%%%%%%%%%%%%%%%%%%%%%%%%%%%%%%
\section{Introduction}
Bolometric luminosity is the total energy per second radiated by an object over all wavelengths in all directions.  Quasars are notable for emitting their substantial radiation over the entire electromagnetic spectrum.  The shape of a quasar spectral energy distributions (SED), see \citet{wilkes04} for a review, depends on the physical properties and structure of the quasar.  Specifically, these are the properties of the central supermassive black hole, the surrounding accretion disc (see \citealt{koratkar99} for a review of emission from this region), the existence of a dusty torus and its geometry, dust heated by starburst regions in the host galaxy \citep{netzer07}, and any radio structures that may exist.  These physical structures differentiate themselves by radiating in different wavelength bands.  In the radio, quasars look like kiloparsec-scale relativistic jets that can extend spatially beyond the host galaxy; in the infrared Ð dust of various temperatures from a parsec-scale torus; and in the optical/ultraviolet Ð sub-parsec-scale accretion disks.  As a result, bolometric luminosity and black hole accretion rate (expressed as $L_{bol}/L_{edd}$), both fundamental quasar properties, are challenging to determine. 

In practice, bolometric luminosity is difficult to measure for several reasons.  Firstly, quasar SEDs cannot be observed, due to a quasar's broad wavelength emission, without data from multiple telescopes.  This makes constructing a full SED expensive in terms of telescope time; a detailed SED can use observations from ten or more telescopes, sometimes using more than one instrument on a single telescope.  

Despite this issue, quasar SEDs have still been compiled.  Using observations from 12 telescopes and 16 instruments, \citet{elvis94} constructed an atlas of 47 quasar SEDs and used them to determine bolometric corrections, which have been used extensively since.  There have been other recent investigations of quasar SEDs as well \citep{shang11,richards06}, those that focus on specific wavelength regimes \citep[e.g.,][]{mullaney11,nb10,netzer07,vasudevan07,shang05}, those that focus on specific classes of objects \citep[e.g.,][]{chen11,gallagher07}, and also those of individual objects \citep[e.g.,][]{zheng01}.  Despite these contributions, the \citet{elvis94} mean SED and bolometric corrections are still often used.

Even when data from multiple telescopes are compiled, there are gaps in the electromagnetic spectrum that remain unobserved.  Gaps in the extreme ultraviolet (EUV) and the near-infrared (NIR) to a lesser extent, are problematic when calculating bolometric luminosity.  The emission in these gap regions has not been observed and can, at best, be modeled \citep[e.g.][]{shakura73,matthews87,korista97} or inferred from spectral properties at other wavelengths \citep[e.g.,][]{kruczek11}.  Another approach is to breach the gap by simply connecting the data on either side with a straight line in log-space \citep[e.g.,][]{elvis94}.

Quasars are variable in many wavelength regions.  Because emission at different wavelengths originates in different physical structures, different spectral regions vary on different timescales.  Optical emission can vary on a timescale of months to years whereas X-ray emission varies on a shorter timescale \citep[][and references therein]{hook94}.  Building a consistent, individual SED requires at least quasi-simultaneous observations on these timescales in spectral regions that suffer from variability.

Quasar emission can be significantly contaminated by light from the host galaxy, particularly in the near and far-infrared.  The NIR is important for bolometric luminosity calculations and, while emission from the quasar usually dominates, the NIR sees a drop in quasar emission and a peak in emission from the host galaxy.  In order to study emission intrinsic to the quasar, SEDs must be corrected for host galaxy contamination.  This is usually done \citep[e.g.,][]{ shang11}, but does add a source of uncertainty to bolometric luminosity calculations.

Quasar emission is anisotropic \citep[e.g.,][]{nb10}.  Anisotropy may vary from object to object and can depend on wavelength.  An improved understanding of disc anisotropy can lead to a correction for average viewing angle to the optical/UV continuum.  \citet{nb10} use the theoretical accretion disc models of \citet{hubeny00} to derive bolometric corrections and explore their dependencies on accretion rate, black hole mass, and viewing angle.  \citet{hubeny00} model the accretion disc in the range $10^{14}-10^{17}$ Hz as a thin disc that includes geometric and relativistic effects.  Due to the lack of any contribution from an X-ray corona, these models do not provide adequate emission at X-ray energies as compared to observed SEDs.  Despite the shortcomings of accretion disc models, they are the best guide for the kinds of corrections that may be necessary as a result of assuming isotropy.

These problems have been overcome to varying extents to allow calculations of bolometric luminosity to be made.  In common practice, bolometric luminosity is estimated by scaling from a monochromatic luminosity.  Different scaling factors exist, e.g. at 5100 \AA: $L_{iso}=13.2\, \lambda L_{\lambda}$ \citep{elvis94}, $L_{iso}=9\, \lambda L_{\lambda}$ \citep{kaspi00}, $L_{iso}=1.5\times9\, \lambda L_{\lambda}$ \citep{shang05}, and $L_{iso}=10.33\, \lambda L_{\lambda}$ \citep{richards06}.  Note that we differentiate between the bolometric luminosity calculated under the assumption of isotropy $(L_{iso})$ and true bolometric luminosity ($L_{bol}$), which likely differ.

There is a need for new bolometric corrections based on up-to-date SEDs with a new methodology.  Pointed observations \citep[e.g. ][]{elvis94} give more detailed individual SEDs than those built from survey data \citep{richards06}, though sample sizes are typically smaller.  However, the \citet{elvis94} SEDs require updating and those built from more recent observations \citep[e.g. ][]{netzer07, shang05} lack broad wavelength coverage.  In addition, the common assumption of isotropy is not a good one.  

The 85 SEDs of \citet{shang11}, intended as an update to those of \citet{elvis94}, are based on pointed observations from the current generation of telescopes.  These SEDs combined with a new methodology provide an opportunity to handle the above issues in a thoughtful, consistent way to calculate the next generation of bolometric corrections.

This paper is organized as follows.  Section 2 describes the data and the \citet{shang11} SEDs.  Section 3 contains a description of the bolometric luminosity calculations.  We derive bolometric corrections in Section 4 and discuss them in the context of previous work.  In Section 5 we calculate X-ray bolometric corrections.  Section 6 summarizes this investigation.

We adopt a cosmology with $H_0 = 70$ km s$^{-1}$ Mpc$^{-1}$, $\Omega_{\Lambda} = 0.7$, and $\Omega_{m} = 0.3$.

%DATA
%%%%%%%%%%%%%%%%%%%%%%%%%%%%%%%%%%%%%%%%%%%%%%%%%%%%%%%%%%%%%%%%%%%%%%%%%%%%%%%%%
\section{Data and Sample Selection}

\subsection{The Full SED Sample}
We derive empirical bolometric corrections from the NIR-to-X-ray continua of 63/85 of the \citet{shang11} SEDs.  This atlas has a total of 85 objects from three different subsamples which are described briefly below.

\begin{itemize}
\item The `PGX' subsample contains 22 of 23 Palomar-Green (PG) quasars in the complete sample selected by \citet{laor94,laor97} to study the soft-X-ray regime.  This subsample is UV bright and has  $z \le 0.4$.  The optical-UV region is covered by UV spectra from Hubble Space Telescope (\emph{HST}) and quasi-simultaneous ground-based optical spectra from McDonald Observatory. \\

\item The `FUSE-HST' subsample has 24 objects, 17 of which come from the Far Ultraviolet Spectroscopic Explorer (\emph{FUSE}) AGN program \citep{kriss00}.  This is a heterogeneous, UV-bright sample with $z<0.5$.  The SEDs for this subsample have quasi-simultaneous \emph{FUSE} \citep{moos00}, \emph{HST}, and Kitt Peak National Observatory (KPNO) observations. \\

\item The `RLQ' subsample includes nearly 50 quasars originally assembled to study orientation; all members of the sample have similar extended radio luminosity which is thought to be isotropic.  The SEDs have quasi-simultaneous \emph{HST} and McDonald or KPNO observations.  See \citet{wills95} and \citet{netzer95} for additional details on this subsample. The blazars originally included in this sample are excluded in the SED atlas because of their variability due to synchrotron emission from a beamed jet. \\
\end{itemize}

X-ray coverage comes from \emph{Chandra}, \emph{XMM}, and \emph{ROSAT} observations that are available in the literature.  When data were available from multiple sources, \emph{Chandra} and \emph{XMM} are given preference over \emph{ROSAT} due to their broader energy coverage and higher sensitivity.  NIR coverage comes from 2MASS photometry \citep{skrutskie06}.  The 2MASS point source catalog has 79 objects in the sample.  The sample also has FIR and radio coverage, though those data are not used in this study.

We used the NIR through X-ray (rest-frame 1$\mu$m to 8 keV) region of the SED for calculating bolometric luminosity.  The data in this region of the SED were collected between 1991 and 2007.  AGN variability, which can be a concern at these wavelengths, does not appear to be a significant issue in these SEDs.  Optical-FUV data were taken quasi-simultaneously (within weeks) and near-to-mid-infrared emission, which arises from a size scale on the order of parsecs \citep[e.g.,][]{raban09} or less for the hottest dust, will vary on a longer timescale of months to years.  We do not see any obvious discontinuities in emission that might indicate variability.

\subsection{Corrections}
In order to assure that the calculated bolometric luminosity of the quasar does not include emission from the host galaxy, it is important to remove any contaminating emission from the host.  The effects of local intervening Galactic gas must also be accounted for.  In the process of constructing the SEDs, \citet{shang11} applied two corrections to the data.  

First, at optical-to-FUV wavelengths the SEDs suffer from Galactic extinction from dust.  This is removed with the empirical mean extinction law of \citet{ccm89} using the dust maps of \citet{schlegel98}.  

Second, \citet{shang11} corrects for host galaxy contamination at NIR and optical wavelengths, although according to those authors, host galaxy contamination to the AGN light is not large for their UV/optically bright quasars.  The host galaxy contribution is maximized compared to the AGN in the H-band for redshifts near $\sim$0.5 \citep{mcleod95} so H-band photometry is used to determine host fraction.  \citet{shang11} make observations with IRTF or HST or collect them from the literature for 28 objects \citep{mcleod94a,mcleod94b,mcleod01}.  

For objects with IRTF observations, host fraction is determined for the H-band using the procedure of \citet{mcleod94a}.  In this procedure, a standard star is observed directly before or after the target.  The stellar profile, or point-spread function (PSF), is then scaled to the innermost pixel of a one-dimensional surface brightness profile for the quasar-plus-host.  The fraction of the PSF necessary to make the quasar-minus-PSF profile turn over in the center is then subtracted to leave a good estimate of host emission.  \citet{shang11} make a minimum and maximum subtraction of the PSF to estimate an uncertainty in this procedure of a few percent, although we note that this method seems to over-subtract and is less trustworthy than subtractions using HST data.  

For objects with HST observations, two-dimensional fitting is possible.  \citet{shang11} use \textsc{galfit} \citep{peng02} to fit several galaxy models in addition to the PSF.  The resulting best fit is then used.  Uncertainty in this method is a few percent or better.

For objects without H-band observations the 2MASS PSF magnitude is taken to be the magnitude of the AGN only and the aperture magnitude is considered total magnitude of the AGN plus the host.  This method is used for the object shown in Fig. \ref{fig:integration}.

Once host fraction has been determined in the H-band, removing the host contamination is straightforward.  An elliptical galaxy template of NGC 0584 from \citet{dale07}, scaled in the H-band, is then used to determine host fraction in the J and K bands and those magnitudes are corrected accordingly.  Fig. \ref{fig:integration} shows the appropriately scaled elliptical galaxy template that is used to determine host fraction for the J and K bands.  

Host galaxy contamination can also affect emission at long wavelengths of the optical spectra, although to a lesser extent.  \citet{shang11} try to remove host contribution here when extracting the spectra.  They use different aperture sizes to ensure that the host contribution is undetectable in the final spectrum. 

\subsection{The Bolometric Luminosity Sample}
We used a subsample of the SEDs for calculating bolometric luminosities.  Of the total 85 objects in the SED atlas, only 63 have the wavelength coverage appropriate for this study.  The following objects were excluded from the sample due to insufficient X-ray coverage: 3C 175, 3C 288.1, 4C 30.25, 4C 64.15, B2 1351+31, MC2 1146+111, MRK 506, PG 1103-006, PG 1259+593, PG 1534+580, PG 2214+139, PG 2251+113, PKS 2216-03, and TEX 1156+213.  After the above objects were removed from the sample, the following lacked sufficient NIR coverage and were also excluded: 3C 110, 3C 186, 3C 207, 4C 12.40, B2 1555+33, B2 1611+34, MC2 0042+101, and PKS 0859-14.  This leaves 63 objects with full wavelength coverage for calculating bolometric corrections.  Table \ref{tab:sample} lists sample members.

The combined sample consists of 23 radio-quiet (RQ) and 40 radio-loud (RL) quasars.  The RQ quasars are all from either the PGX or FUSE-HST samples and are lower redshift ($z<0.5$), while the RL quasars come primarily from the RLQ sample and more than half have higher redshifts ($z>0.5$).  The RL quasars have an average luminosity about 6 times higher than those that are RQ.  Both RL and RQ quasars span approximately 2 orders of magnitude in luminosity.  While not statistically complete or well-matched, the combined sample is comprised of subsamples with specific selection criteria.  The `PGX' subsample may be complete, although it is not representative of the general quasar population so we advise caution when using corrections derived here on more general samples.  The high flux limit of the PG sample means that these bright objects are also uncharacteristically blue and \citet{jester05} specifically points out that the PG sample is biased against redder spectra.  The `FUSE-HST' subsample was selected randomly among objects known to have high UV fluxes.  The `RLQ' subsample was selected to have similar extended radio luminosity, a quantity thought to be independent of orientation.  In order to find large numbers of these objects it can be necessary to go to higher luminosity and redshift.

\begin{figure}
\begin{center}
\includegraphics[width=8.9 truecm]{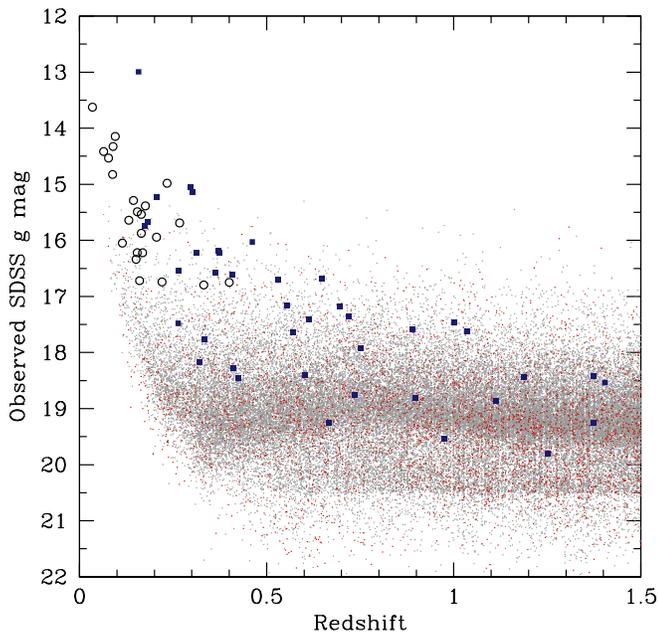}
\end{center}
\caption{The region of observed parameter space occupied by the 63 objects in this sample and the SDSS DR7 quasar catalog.  SDSS quasars are small gray points, SDSS RL quasars are distinguished with small red points, RL quasars from this work are blue squares, and RQ quasars from this work are open black circles.  The bolometric corrections derived here may lose accuracy for objects that occupy another region of parameter space.}
\label{fig:sample}
\end{figure}

We compared our sample in observed parameter space to the Sloan Digital Sky Survey (SDSS) Data Release 7 (DR7) quasar catalog \citep{schneider10}.  Fig. \ref{fig:sample} shows the observed SDSS g magnitude vs. $z$.  We calculated approximate SDSS g magnitude for our objects by converting the flux at the center of the g-band to a magnitude with the SDSS flux zero point.  The radio loudness parameter, log $R^{*}$, is estimated for SDSS objects with FIRST detections at 20 cm using the prescription of \citet{stocke92} (note that we substituted $g$ for $B$).  The bolometric corrections derived here will be most accurate for objects that lie in the region of parameter space occupied by this sample and may be less reliable when used on dissimilar objects.

While the majority of this sample was selected to be bright in the UV, there may be intrinsic reddening in some objects.  It can be difficult to differentiate between intrinsic reddening and an extreme SED \citep[e.g.,][]{zheng01}.  In light of this, we did not exclude objects at this point.

%DETERMINING BOLOMETRIC LUMINOSITY
%%%%%%%%%%%%%%%%%%%%%%%%%%%%%%%%%%%%%%%%%%%%%%%%%%%%%%%%%%%%%%%%%%%%%%%%%%%%%%%%%
\section{Determining Bolometric Luminosity}
To know the bolometric luminosity exactly, it is necessary to observe a quasar at all wavelengths and from all directions.  Since we are restricted to one line of sight and can only observe parts of the electromagnetic spectrum, we must make some assumptions in order to calculate bolometric luminosity.  These assumptions, and our method, are outlined below.

\subsection{Empirical Isotropic Luminosity}
The measured observed frame flux is converted to an observed luminosity under the assumption, which we will re-examine later (\S \ref{sec:isotropy}), that emission is isotropic.

\begin{equation}
L_{obs,\nu_{obs}} =  \int_{0}^{2\pi}\int_{0}^{\pi} F_{obs,\nu_{obs}}\,d_{L}^2 \sin{\theta}\,d\theta\,d\phi
\label{eq:integral}
\end{equation}
\begin{equation}
L_{obs,\nu_{obs}} = 4\pi \, d_{L}^2 \, F_{obs,\nu_{obs}},
\label{eq:result}
\end{equation}

\noindent where $d_{L}$ is the luminosity distance, $F_{obs,\nu_{obs}}$ is the observed flux density, and $\nu_{obs}$ is the observed frequency.  The integration between Equation~\ref{eq:integral} and Equation~\ref{eq:result} that results in the $4\pi$ coefficient is possible only under the assumption of isotropy.  Then the observed luminosity is converted to rest luminosity via the equation:

\begin{equation}
L_{rest,\nu_{rest}} = \frac{L_{obs,\nu_{obs}}}{1+z} =\frac{4 \pi d_{L}^2F_{obs,\nu_{obs}}}{1+z},
\end{equation}

\noindent where $\nu_{rest}$ is the rest frequency.  Bolometric luminosity is then calculated by integrating $L_{rest,\nu_{rest}}$ over frequency, where specific consideration should be given to the limits of integration.

Emission at IR wavelengths is thought to result from optical-UV photons reprocessed by surrounding hot dust \citep[e.g.,][]{barvainis87}.  Under the assumption of isotropy, it is important not to include reprocessed photons.  Assuming isotropy requires that the optical-UV photons that we observe are characteristic of radiation from the accretion disc in all directions.  We count these photons for the first time when we integrate the optical-UV over all space.  IR photons are likely optical-UV photons originally emitted in another direction that have been reprocessed and re-emitted along our line of sight by a AGN-heated dust.  We count these photons a second time if we integrate the IR portion of the SED.

\begin{figure*}
\begin{center}
\includegraphics[width=13.0 truecm, angle=-90]{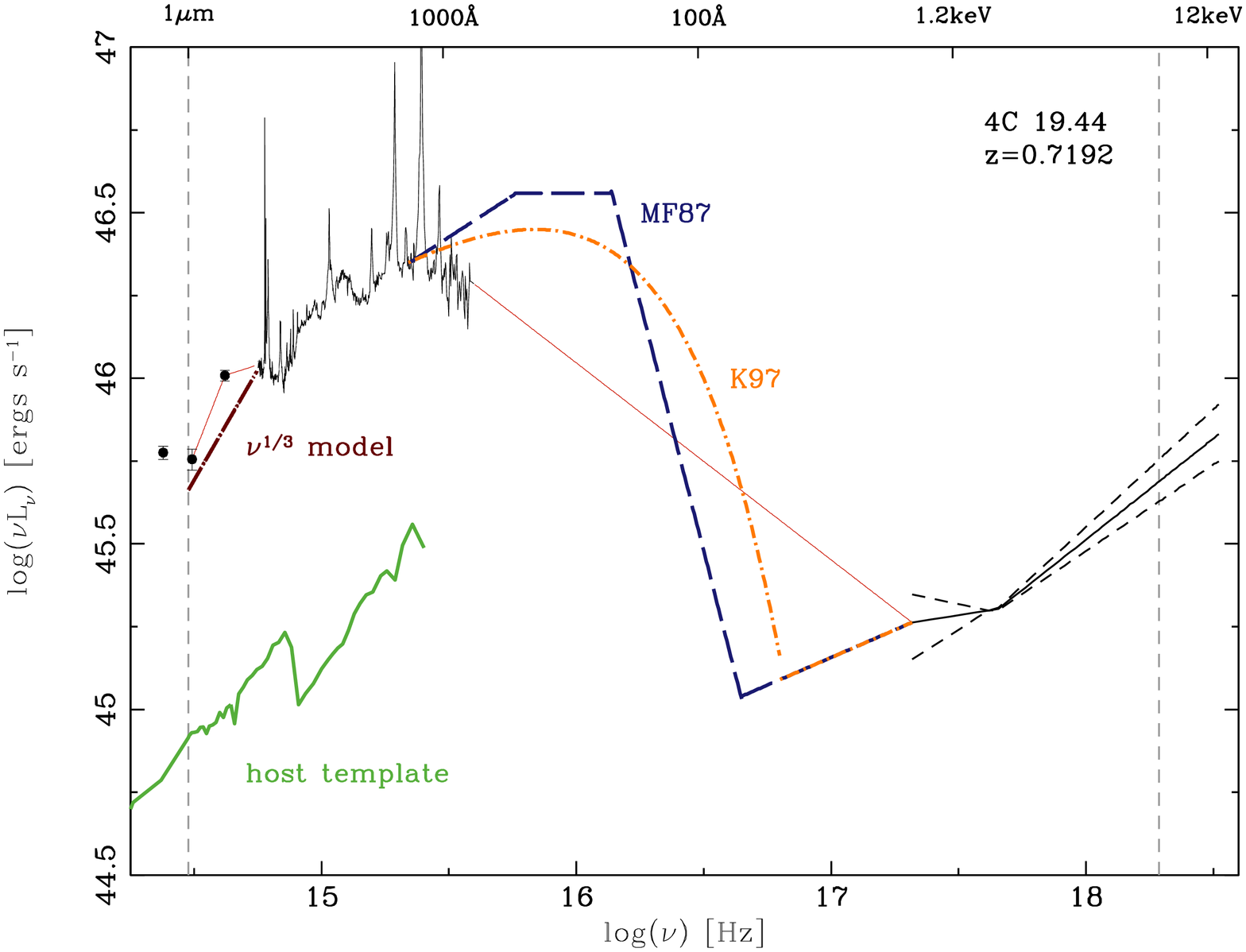}
\end{center}
\caption{Data in the NIR to X-ray region of the SED for 4C 19.44 shown by solid black line with the region of integration denoted by dashed gray vertical lines.  In the NIR the $\nu^{1/3}$ disc model is shown as a dark red, dashed-dotted line.  In the FUV-soft-X-ray, the \citet{matthews87} model is shown as a blue dashed line and the \citet{korista97} is shown as an orange dotted-dashed line.  The power law interpolations that we ultimately employ are shown as solid red lines.  The solid green line shows the elliptical galaxy template for NGC 0584 from \citet{dale07} used in \citet{shang11} for host galaxy subtractions to the 2MASS photometry.  The host template is scaled to 15\%, the host galaxy fraction for this object, of the 2MASS H-band magnitude.  Wavelength and energy are given on the top axis.}
\label{fig:integration}
\end{figure*}

To accurately estimate the bolometric luminosity under the assumption of isotropy, the IR region should be excluded from the integration.  There has not been agreement in the field on this point: some \citep[e.g. ][]{elvis94,richards06} integrate the entire SED whereas others \citep[e.g. ][]{netzer07,nb10} integrate only the region in emission at optical-through-X-ray wavelengths (i.e. the Big Blue Bump).  Here we integrate the SED from 1~$\mu$m to 8 keV ($\nu = 3\times10^{14}-1.93\times10^{18}$Hz).  Note that we assume that there is no significant energy contribution at very high energies ($>8$ keV).  This assumption is made perforce because the data do not extend past this limit.  The region of integration is shown in Fig. \ref{fig:integration} for a typical quasar SED.  Energy output at radio wavelengths is insignificant compared to emission from the accretion disk so it can be excluded without significantly altering the integrated luminosity.

An alternative to this method is to use the integrated mid-infrared (MIR) emission and covering fraction as a measure of bolometric luminosity \citep[e.g.,][]{spinoglio89,edelson86}.  This method has several advantages since the MIR is less susceptible to reddening and likely much more isotropically emitted than the Big Blue Bump.  In particular, such a correction would be valuable for reddened and Type 2 objects.  It would not be without its own issues, however.  Corrections for starlight would be necessary at the shortest wavelengths and also for starburst-heated dust in the host galaxy \citep{netzer07} at the longest wavelengths, depending on the limits of integration.  Additionally, depending on redshift and available facilities, the MIR bump may be more difficult to observe than the Big Blue Bump and also has gaps in data coverage, although they may be easier to fill.  Finally, while the MIR emission may be more isotropically emitted, the anisotropy of the optical/UV emission that acts as a source for MIR reprocessing must be accounted for.  
 
\subsection{Integrating in the NIR}
In order to integrate the SED, the NIR region around log$(\nu) \sim 14.25-14.75$ required special consideration.  Data coverage here was patchier than the optical/UV (2MASS photometry versus a spectrum) and in some objects the 2MASS data did not reach the 1 $\mu$m limit of integration in the rest frame.

In the NIR we used a power law interpolation between 2MASS data points and then, when necessary, extrapolated out from the longest wavelength 2MASS point to provide data all the way to the 1 $\mu$m limit of integration.  Note that, because we chose a hard 1 $\mu$m rest frame limit of integration but had a range of redshifts, we integrated different data in each SED (e.g. in the object shown in Figure~\ref{fig:integration} we did not integrate between the H and K bands or beyond because the 1 $\mu$m limit had already been reached).

In order to quantify interpolating between the 2MASS points and extrapolating beyond, we replaced these interpolation/extrapolation power laws with the $F_{\nu} \propto \nu^{1/3}$ \citep{shakura73} accretion disc model extrapolated out from the optical spectrum to 1 $\mu$m and recalculate bolometric luminosity.  The $\nu^{1/3}$ model is reached by integrating a multiple blackbody accretion disc over infinite radius and is appropriate for the long wavelength emission of the disc (i.e. the NIR).  Both the $\nu^{1/3}$ and power law interpolation models are shown in Fig. \ref{fig:integration}.  

\citet{kishimoto05} find good agreement between the $\nu^{1/3}$ model and the observed emission from quasar accretion discs.  Their result is primarily based on PG 1425+267, an object in our sample, as well as two other broad line quasars of similar redshift to our sample (several tenths).  

We normalized the $\nu^{1/3}$ model between 5000 and 5300 \AA, a region relatively free of emission lines, and which should have little contribution from hot dust emission.  Emission at these wavelengths may be contaminated by residual host galaxy emission, even after the correction of \citet{shang11}.  We may have introduced some small uncertainty as a result, but this is not larger than is introduced by implementing the $\nu^{1/3}$ disc model.

We found that the difference in bolometric luminosity between our power law interpolation and the $\nu^{1/3}$ disc model is small, on average $-2\pm2$ per cent.  Bolometric luminosity resulting from $\nu^{1/3}$ model is plotted against bolometric luminosity resulting from the power law interpolation in Figure~\ref{fig:interps}.

\subsection{Integrating in the X-ray}
The X-ray region of the SED also required special measures to be taken before it could be integrated.  There were two issues in this region: in some objects we did not have data out to our 8 keV limit of integration and there was a large gap in data coverage for all objects between the FUV and X-ray around log$(\nu) \sim 15.75-17.25$.

In some objects we had to extend the X-ray power law to higher energies to reach the 8 keV limit of integration.  Observations in this region were a magnitude or more down from the peak of the SED, so changing this limit did not significantly change the integrated luminosity.

Our region of interest in the SEDs contained a large gap in data coverage: one between the FUV and X-ray.  We bridged this gap with a linear interpolation in log $\nu L_{\nu}$ space.  This power law appeares to be a good description of this region in some cases \citep[e.g.,][]{laor97}.  Fig. \ref{fig:integration} shows the power law interpolation as well as models from \citet{matthews87} and \citet{korista97}, described below, over-plotted for one representative object.

While a straight-forward power law did seem to connect FUV and X-ray data well in some cases, we recalculated bolometric luminosity using two other models to help quantify the level of uncertainty that can result from selecting different models in the EUV.

Both models were normalized to the data near 1300 \AA, a region also relatively free of emission lines.  In cases where the model did not intersect the X-ray data, the X-ray data were extended to longer wavelengths until they intersected the model. 

The first model is the SED of \citet{matthews87}, which they derive from continuum observations when available and otherwise infer from the emission-line spectrum.  \citet{matthews87} model the region between the FUV and X-ray with several power laws.  

The second model is that of \citet{korista97}.  They model the UV region with a modified power law of the form $F_{\nu} \propto \nu^{-0.5}e^{-h\nu/kT_{cut}}$ (a thermal bump). At 912 \AA\ and shorter wavelengths, the power law $F_{\nu}\propto \nu^{-1}$ is used.  We used the SED quoted by those authors as typical for an average quasar: $T_{cut} = 10^6$ K, $E_{peak} = 44$ eV.

The percent difference in bolometric luminosity caused by using the \citet{matthews87} model instead of a power law interpolation is $41\pm21$, where the given uncertainty is the standard deviation.  Using the \citet{korista97} model caused a percent difference in bolometric luminosity of $33\pm20$.  Visual inspection showed that in many cases, the \citet{matthews87} and \citet{korista97} models may not be a good description of the SED; in some objects these models over or under-estimated the X-ray data by as much as an order of magnitude for over an order of magnitude in frequency.  In a few objects these models always over-estimated the data by approximately 0.1 dex.

\begin{figure*}
\begin{minipage}[!b]{5.25cm}
\centering
\includegraphics[width=6cm]{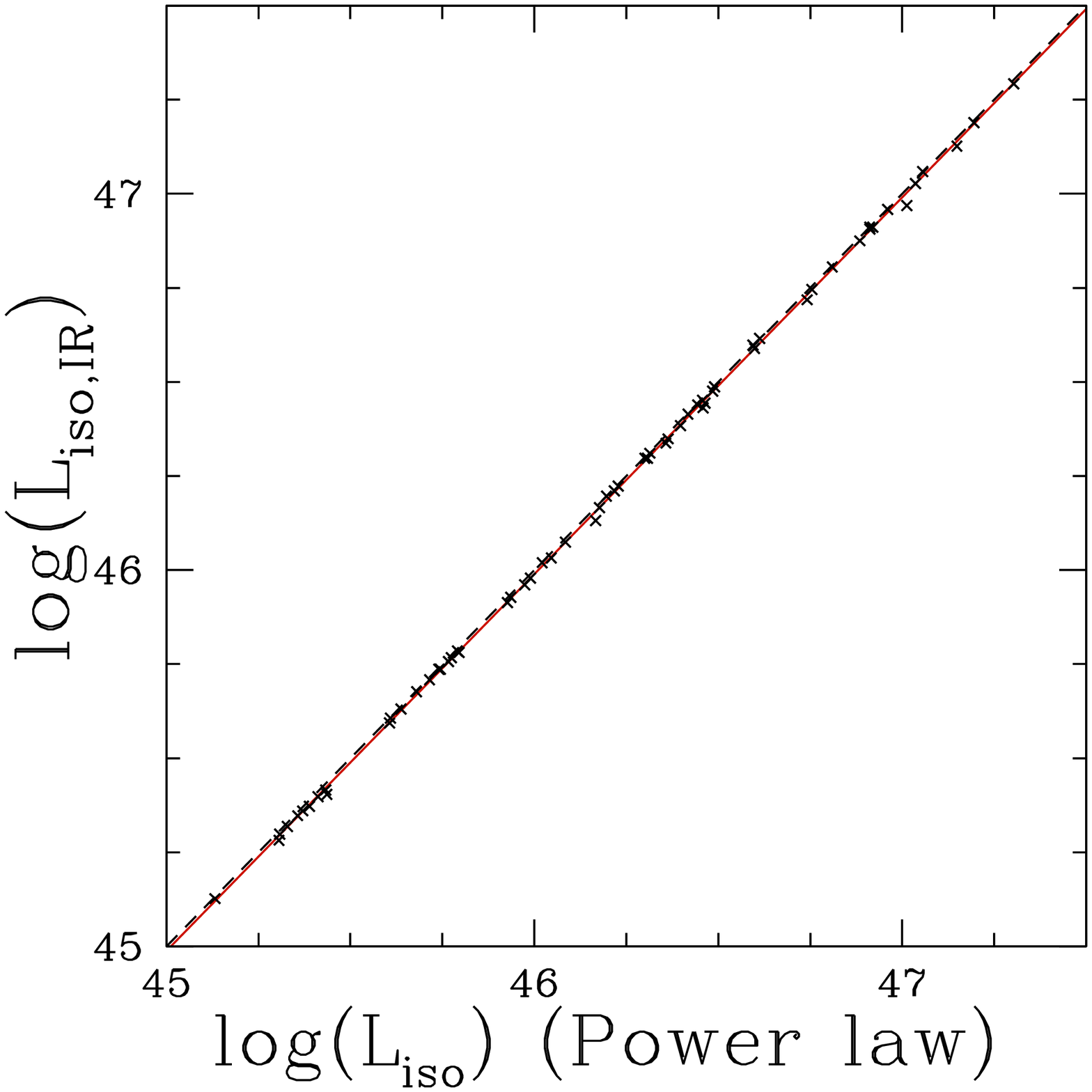}
\end{minipage}
\hspace{0.6cm}
\begin{minipage}[!b]{5.25cm}
\centering
\includegraphics[width=6cm]{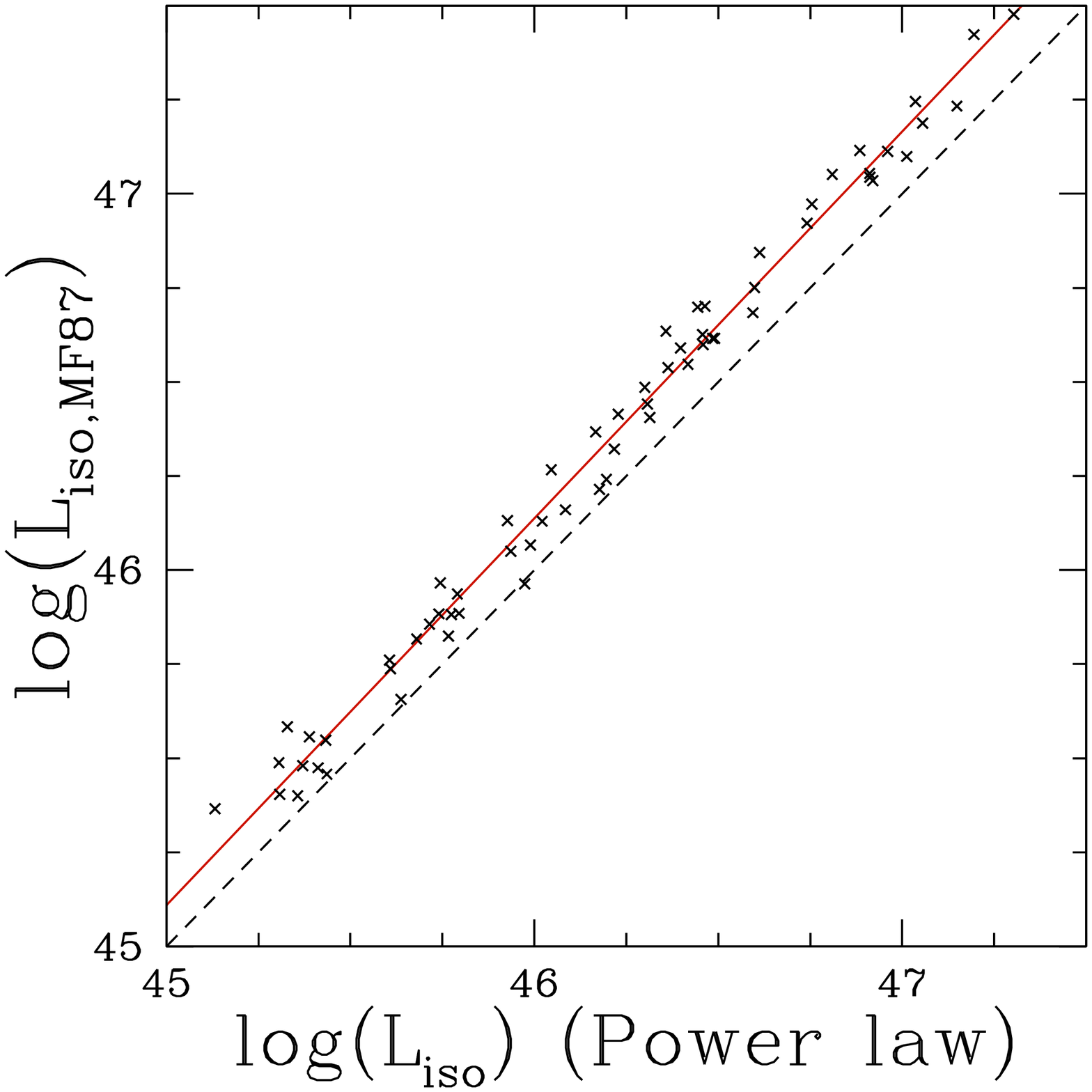}
\end{minipage}
\hspace{0.6cm}
\begin{minipage}[!b]{5.25cm}
\centering
\includegraphics[width=6cm]{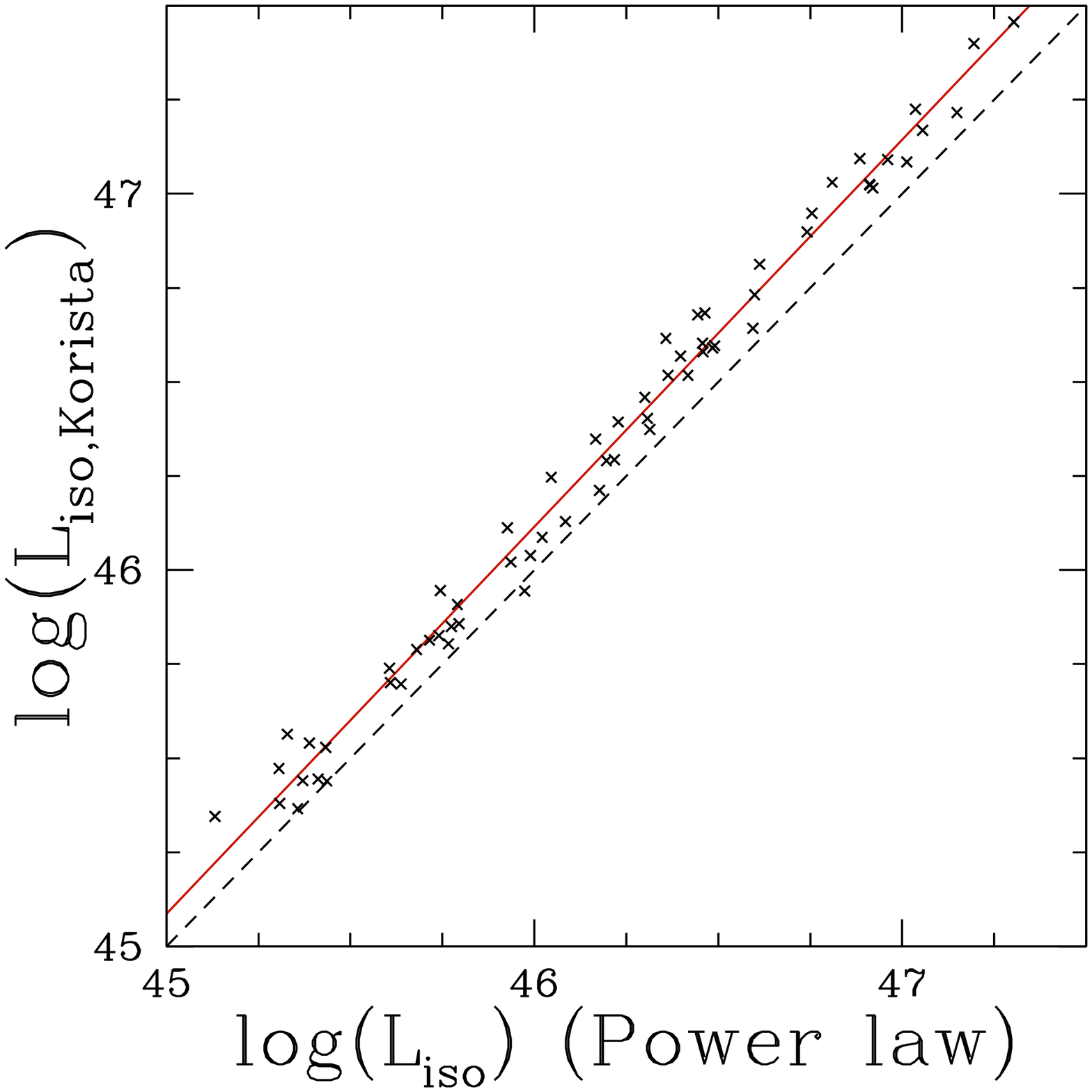}
\end{minipage}
\caption{Bolometric luminosities resulting from $\nu^{1/3}$, \citet{matthews87}, and \citet{korista97} interpolation methods plotted against bolometric luminosities resulting from the direct power law interpolation that is ultimately employed.  Solid red lines are best-fitting models to the data, black dashed lines are one-to-one lines. \label{fig:interps}}
\end{figure*}

Bolometric luminosities resulting from the \citet{matthews87} and \citet{korista97} models are each plotted against bolometric luminosities resulting from the power law interpolation in Fig \ref{fig:interps}.  This figure illustrates how significantly our ignorance of emission in this FUV-X-ray gap impacts our bolometric luminosities; \citet{matthews87} and \citet{korista97} X-ray models tend to result in much larger bolometric luminosities than a power law interpolation with lots of variation between objects.

It has been suggested that the intrinsic SED continuum may be different from what we view \citep[e.g.,][]{korista97,lawrence11} and some tentative evidence may exist to support this \citep[e.g.,][]{richards11,kruczek11}.  If this is the case, this approach is still at least energetically viable.  Whether the energy holds the same shape as when it was originally emitted from the accretion disc or not, it has still been emitted and should still be counted towards the bolometric luminosity.  We note one caveat to this: if there is a large peak in emission that is unobserved in the EUV it will go unaccounted for and our calculation of bolometric luminosity will underestimate the true value as a result.  If emission is reprocessed in the same wavelength regime, double counting will not be an issue.  Uncertainty will still come from how isotropically different physical regions emit.

Integrated luminosities assuming isotropy ($L_{iso}$) are presented in Table \ref{tab:sample}.  Radio-loudness, given in that table, is calculated by \citet{shang11} from R* = f(5 GHz)/f(4215 \AA).

\begin{table*}
\begin{minipage}{13cm}
\caption{Physical Data \label{tab:sample}}
\scalebox{0.9}{
\begin{tabular}{lccccccccccc}
\hline
Object & Redshift &  log(R*) \footnote{Radio loudness, R = f(5 GHz)/f(4215 \AA)} & $\alpha_{\lambda,opt}$ & log $(L_{iso})$ \footnote{Integrated luminosity assuming isotropy.} &  log $(L_{bol})$ \footnote{Integrated luminosity adjusted for viewing angle: $L_{bol}\approx0.75\,L_{iso}$} & $\zeta_{5100}$ & $\zeta_{3000}$ & $\zeta_{1450}$ \\
\hline	
3C 215			&   0.4108      	&  	3.37	&  	-0.69		&	  45.77	&   45.64  &    	6.77      	&	  6.66         &	  5.54            	\\
3C 232			&   0.5297        	&  	2.87	&    	-1.55		&	  46.36	&   46.23  &    	4.30      	&	  2.64         &	  3.06            	\\
3C 254			&   0.7363        	&  	3.71	&    	-1.41		&	  46.17	&   46.04  &    	6.37      	&	  4.26         &	  3.47            	\\
3C 263			&   0.6464        	&  	3.00	&    	-1.96		&	  46.81	&   46.68  &    	6.33      	&	  4.00         &	  3.00            	\\
3C 277.1			&   0.3199     	&  	3.52	&       -1.51	&	  45.61	&   45.48  &    	8.74 		&	  5.73      	&     	  3.69        		\\
3C 281			&   0.6017        	&  	3.23	&    	-1.00		&	  46.23	&   46.10  &    	8.04      	&	  5.05         & 	  3.54            	\\
3C 334			&   0.5553        	&   	3.11	&    	-1.75		&	  46.44	&   46.32  &    	7.30      	&	  3.79         & 	  2.79            	\\
3C 37			&   0.6661          &    	3.74	&  	-3.65		&	  46.18 	&   46.05  & 	19.54         &	  8.82         &	  7.86            	\\
3C 446			&   1.4040        	&    	4.34	&	\nodata	&	  47.01	&   46.89  &    	3.81       	&	  4.92         & 	  6.58            	\\
3C 47			&   0.4250          &    	3.82	&  	-1.22		&	  45.97 	&   45.85  &  	12.31         &	  8.27         &	  5.05            	\\
4C 01.04			&   0.2634     	&     	3.41	&    	 0.15		&	  45.44	&   45.31  &    	5.60 		&	  6.39      	&     	  10.1         	\\
4C 06.69			&   1.0002     	&     	3.32	&     	-0.97		&	  47.30	&   47.18  &    	7.44 		&	  5.84       	&    	  3.71        		\\
4C 10.06			&   0.4075     	&     	2.67	&       -1.74	&	  46.48	&   46.36 &  	10.26  	&	  6.01      	&     	  4.34        		\\
4C 11.69			&   1.0370     	&     	3.78	& 	\nodata	&   	  47.04	&   46.91  &    	4.53 		&	  4.18       	&    	  3.37        		\\
4C 19.44			&   0.7192     	&     	3.42	&     	-0.53		&	  46.91	&   46.79  &    	8.30 		&	  5.67      	&     	  4.11        		\\
4C 20.24			&   1.1135     	&     	3.62	&       -1.41	&	  46.92	&   46.79  &    	9.23 		&	  6.37       	&    	  5.04        		\\
4C 22.26			&    0.9760     	&     	3.26	&        0.42		&	  46.59	&   46.47 &  	11.94    	&	  8.68      	&    	  5.86        		\\
4C 31.63			&    0.2952     	&     	2.93	&       -2.18	&	  46.61	&   46.49	&  	10.14    	&	  4.17      	&     	  2.54        		\\
4C 34.47			&    0.2055     	&     	2.62	&       -1.94	&	  46.08	&   45.96	&  	12.92    	&	  5.98      	&     	  6.47        		\\
4C 39.25			&    0.6946     	&     	3.65	&       -1.74	&	  46.91	&   46.79 	&    	8.76 		&	  6.14      	&     	  4.54        		\\
4C 40.24			&    1.2520     	&     	3.94	& 	\nodata	&	  46.60	&   46.47 	&    	7.24 		&	  5.82       	&    	  4.32        		\\
4C 41.21			&    0.6124     	&     	2.91	&       -1.77	&	  46.75	&   46.63	&  	13.01    	&	  6.35      	&     	  3.03        		\\
4C 49.22			&    0.3333     	&     	3.35	&       -1.00	&	  45.99	&   45.86	&  	14.12    	&	  8.07      	&     	  6.04        		\\
4C 55.17			&    0.8990     	&     	3.74	&     	-0.45		&	  46.46	&   46.33 	&    	5.54 		&	  4.95      	&     	  4.21        		\\
4C 58.29			&    1.3740    	&     	2.65	& 	\nodata	&	  47.19	&   47.07 	&    	5.98 		&	  4.56       	&    	  2.89        		\\
4C73.18			&    0.3027     	&     	3.20	&     	-0.77		&	  46.40	&   46.27  	&    	7.09 		&	  3.86      	&     	  3.95        		\\
B2 0742			&    0.4616     	&     	2.77	&       -1.73	&	  46.46	&   46.34 	&    	4.91  	&	  3.23        	&    	  3.76         	\\
IRAS F07546+3928	&    0.0953 	&     	-0.6	&       -1.38	&	  45.43	&   45.31  	&    	3.75		&	  2.79	&     	  4.72  		\\
MRK 509			&    0.0345     	&	-0.2	&      	-1.64		&	  45.36	&   45.23  	&     	14.74      	&	  7.86        	&    	  6.97         	\\
OS 562			&    0.7506        	&  	3.35	&  	-0.81		&	  46.74	&   46.62 &    	8.31      	&	  5.80         & 	  3.41         	\\
PG  0052+251		&    0.1544     	& 	-0.4	&       -1.73	&	  45.94	&   45.81  	&      	11.36     	&	  6.35        	&    	  4.60         	\\
PG  0844+349		&    0.0643     	&   	-1.1	&      	-1.02		&	  45.31	&   45.18  	&       8.35   	&	  4.91        	&    	  4.50         	\\
PG  0947+396		&    0.2057     	&	-0.5	&      	-1.29		&	  45.79	&   45.67 	&      	10.14     	&	  5.42        	&    	  4.42         	\\
PG  0953+414		&    0.2338     	&	-0.2	&      	-2.16		&	  46.42	&   46.29 	&      	10.80      	&	  5.23        	&    	  3.50         	\\
PG  1001+054		&    0.1603     	&	0.05	&       -2.12	&	  45.13	&   45.01 	&      	5.37   	&	  3.30        	&    	  2.98         	\\
PG  1100+772		&    0.3114     	&    	2.65	&       -2.04	&	  46.46	&   46.33 	&      	10.45      	&	  6.16        	&    	  3.64         	\\
PG  1114+445		&    0.1440     	& 	-0.9	&       -1.50	&	  45.39	&   45.26 	&       3.91   	&	  2.82        	&    	  4.34         	\\
PG 1115+407		&    0.1541     	&	-0.4	&      	-1.36		&	  45.71	&   45.59 	&      	12.82      	&	  6.69        	&    	  4.06         	\\
PG 1116+215		&    0.1759     	&	-0.1	&      	-2.11		&	  46.31	&   46.19 	&      	16.42      	&	  7.86        	&    	  4.25         	\\
PG 1202+281		&    0.1651     	& 	0.04	&       -2.77	&	  45.41	&   45.29 	&       8.22   	&	  5.19        	&    	  8.59         	\\
PG 1216+069		&    0.3319     	&  	0.67 &       -1.91		&	  46.31	&   46.18 	&      	6.66   	&	  6.44        	&    	  4.89         	\\
PG 1226+023		&    0.1576     	&    	3.22	&       -2.00	&	  46.96	&   46.83 	&      	9.88   	&	  5.05        	&    	  3.23         	\\
PG 1309+355		&    0.1823     	&    	1.38	&       -1.38	&	  45.74	&   45.62 	&      	4.73   	&	  3.55        	&    	  3.42         	\\
PG 1322+659		&    0.1684     	&	-0.7	&      	-1.86		&	  45.74	&   45.61 	&      	11.06      	&	  6.66        	&    	  4.64         	\\
PG 1351+640		&    0.0882  	& 	0.09	&       -1.19	&	  45.30	&   45.18 	&       2.55   	&	  3.31       	&     	  3.99         	\\
PG 1352+183		&    0.1510     	&	-0.6	&      	-1.84		&	  45.61	&   45.48 	&      	12.32      	&	  6.40        	&    	  4.44         	\\
PG 1402+261		&    0.1650     	&	-0.5	&      	-2.14		&	  46.02	&   45.90 	&      	15.28      	&	  6.54        	&    	  4.12         	\\
PG 1411+442		&    0.0895    	& 	-0.8	&      	-1.80		&	  45.33	&   45.20 	&       4.57   	&	  2.77        	&    	  2.96         	\\
PG 1415+451		&    0.1143     	&	-0.5	&      	-1.08		&	  45.37	&   45.24 	&      	8.25   	&	  6.03        	&    	  4.63         	\\
PG 1425+267		&    0.3637     	&     	2.31	&       -1.84	&	  46.05	&   45.92 	&       5.88   	&	  3.33        	&    	  3.21         	\\
PG 1427+480		&    0.2203     	&  	-1.5	&      	-1.82		&	  45.79	&   45.66 	&      	11.24      	&	  5.91        	&    	  3.78         	\\
PG 1440+356		&    0.0773    	& 	-0.7	&       -1.44	&	  45.64	&   45.51 	&       10.91     	&	  6.45        	&    	  4.68         	\\
PG 1444+407		&    0.2673     	& 	-1.0	&      	-2.17		&	  46.19	&   46.07 	&      	10.71      	&	  4.85        	&    	  4.41         	\\
PG 1512+370		&    0.3700     	&     	2.85	&       -1.97	&	  46.22	&   46.09 	&       11.57     	&	  4.88        	&    	  4.62         	\\
PG 1543+489		&    0.4000     	&  	0.13	&       -2.15	&	  46.30	&   46.17 	&      	8.33   	&	  4.49        	&    	  3.15         	\\
PG 1545+210		&    0.2642     	&     	3.00	&       -1.69	&	  45.93	&   45.80 	&       7.178   	&	  4.39        	&    	  3.46         	\\
PG 1626+554		&    0.1317     	&  	-1.0	&      	-1.92		&	  45.68	&   45.55 	&      	11.70      	&	  5.94        	&    	  4.15         	\\
PG 1704+608		&    0.3730     	&     	2.82	&       -1.05	&	  46.49	&   46.36 	&    	5.89  	&	  4.78        	&    	  3.82         	\\
PG 2349$-$014	&    0.1740     	&     	2.74	&       -1.48	&	  45.77	&   45.65 	&    	9.98     	&	  5.56        	&    	  4.80         	\\
PKS 0112$-$017	&   1.3743    	&     	3.45	&  	\nodata	&	  46.88	&   46.76 	&     	6.46 		&	  4.41      	&     	  2.80       		\\
PKS 0403$-$13	&    0.5700    	&      	3.73	&      	-0.94		&	  46.36	&   46.24 	&       6.78 		&	  5.24     	&      	  4.67       		\\
PKS 1127$-$14	&    1.1870   	&      	3.88	&       -1.88	&	  47.15	&   47.02 	&       8.70 		&	  6.29      	&     	  4.50       		\\
PKS 1656+053          &    0.8890         &      3.10	&       -1.09	&         47.05	&   46.93 	&	6.29 		&	  5.51   	&      	  4.12     		\\
\hline
\end{tabular}
}
\end{minipage}
\end{table*}

%FORMULATING BOLOMETRIC CORRECTIONS
%%%%%%%%%%%%%%%%%%%%%%%%%%%%%%%%%%%%%%%%%%%%%%%%%%%%%%%%%%%%%%%%%%%%%%%%%%%%%%%%%
\section{Deriving Bolometric Corrections}
Given measurements of a monochromatic luminosity and our estimates of integrated luminosity, it is possible to explore a variety of bolometric corrections.  We made measurements of monochromatic luminosity at three commonly used wavelengths, 1450, 3000, and 5100 \AA.  These wavelengths correspond to the optical window for low, medium, and high-redshift objects, respectively.  When deriving optical/UV corrections we did not distinguish between RL and RQ objects, which we justify in section \ref{sec:radioclass}.  For bolometric corrections to the 2-10 keV luminosity see section \ref{sec:xray}.

\subsection{Linear Bolometric Corrections}
Historically, bolometric corrections assume a linear relationship between bolometric and a monochromatic luminosity:

\begin{equation}
L_{iso} = \zeta \, \lambda L_{\lambda},
\end{equation}

\noindent where $L_{\lambda}$ is the monochromatic luminosity and the factor $\zeta$ is the bolometric correction. Using a bolometric correction of this form assumes a single SED shape scaled to match a monochromatic luminosity.  Quasar SEDs vary in shape and there is no reason to assume a direct linear relationship with a zero intercept, although available data has rarely supported more detailed fittings.  

\begin{figure*}
\begin{minipage}[!b]{5cm}
\centering
\includegraphics[width=6cm]{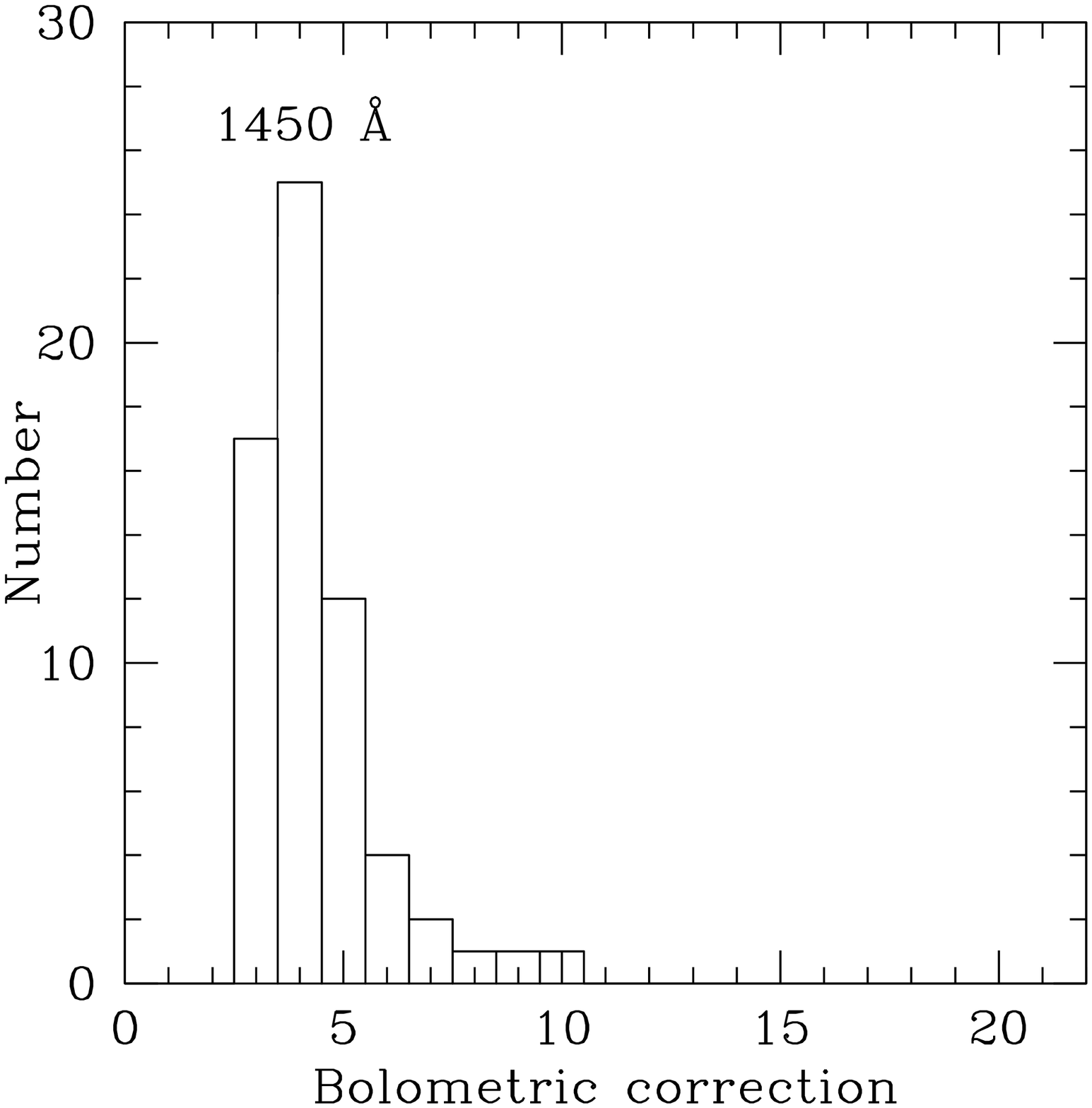}
\end{minipage}
\hspace{0.6cm}
\begin{minipage}[!b]{5cm}
\centering
\includegraphics[width=6cm]{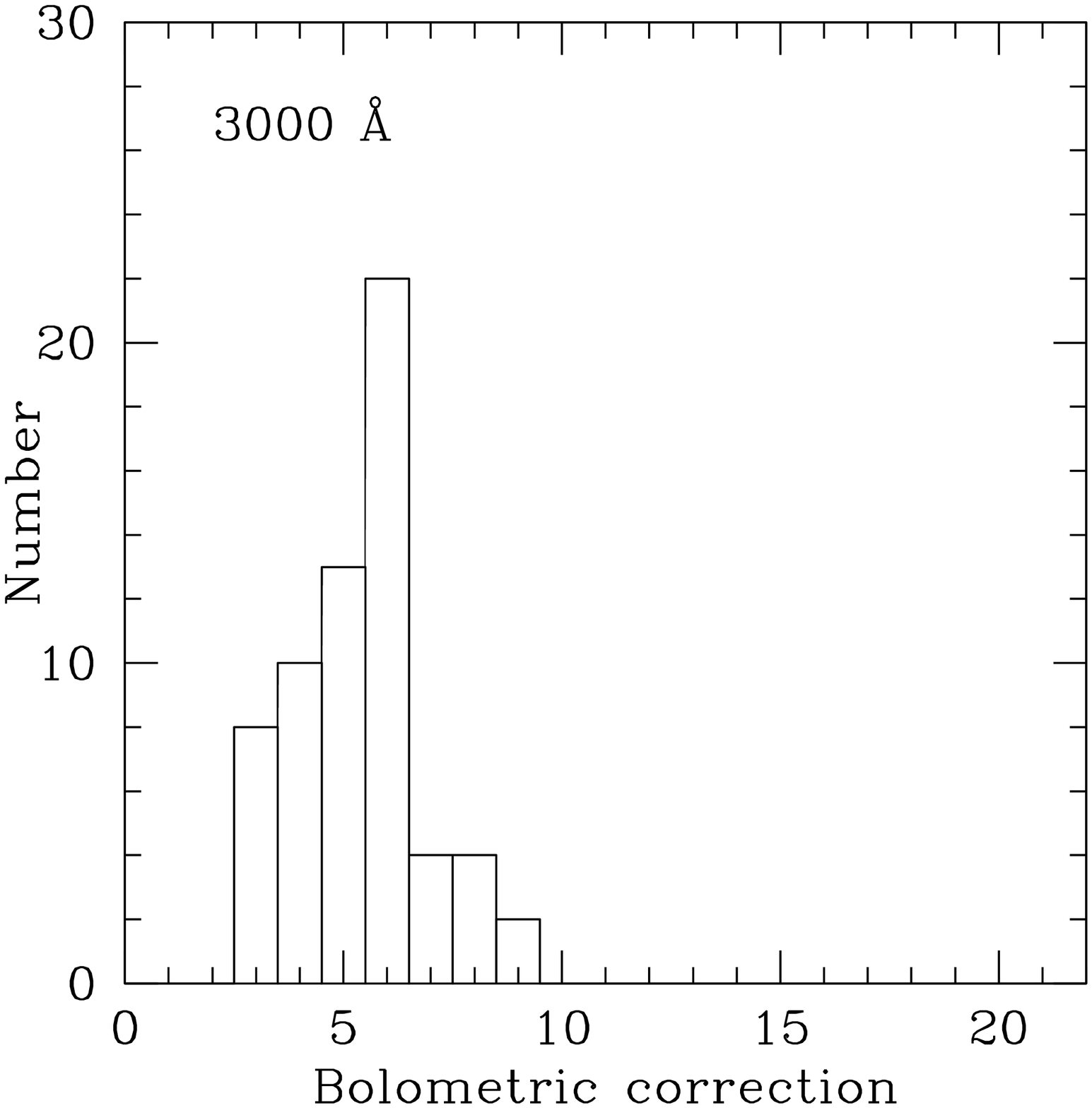}
\end{minipage}
\hspace{0.6cm}
\begin{minipage}[!b]{5cm}
\centering
\includegraphics[width=6cm]{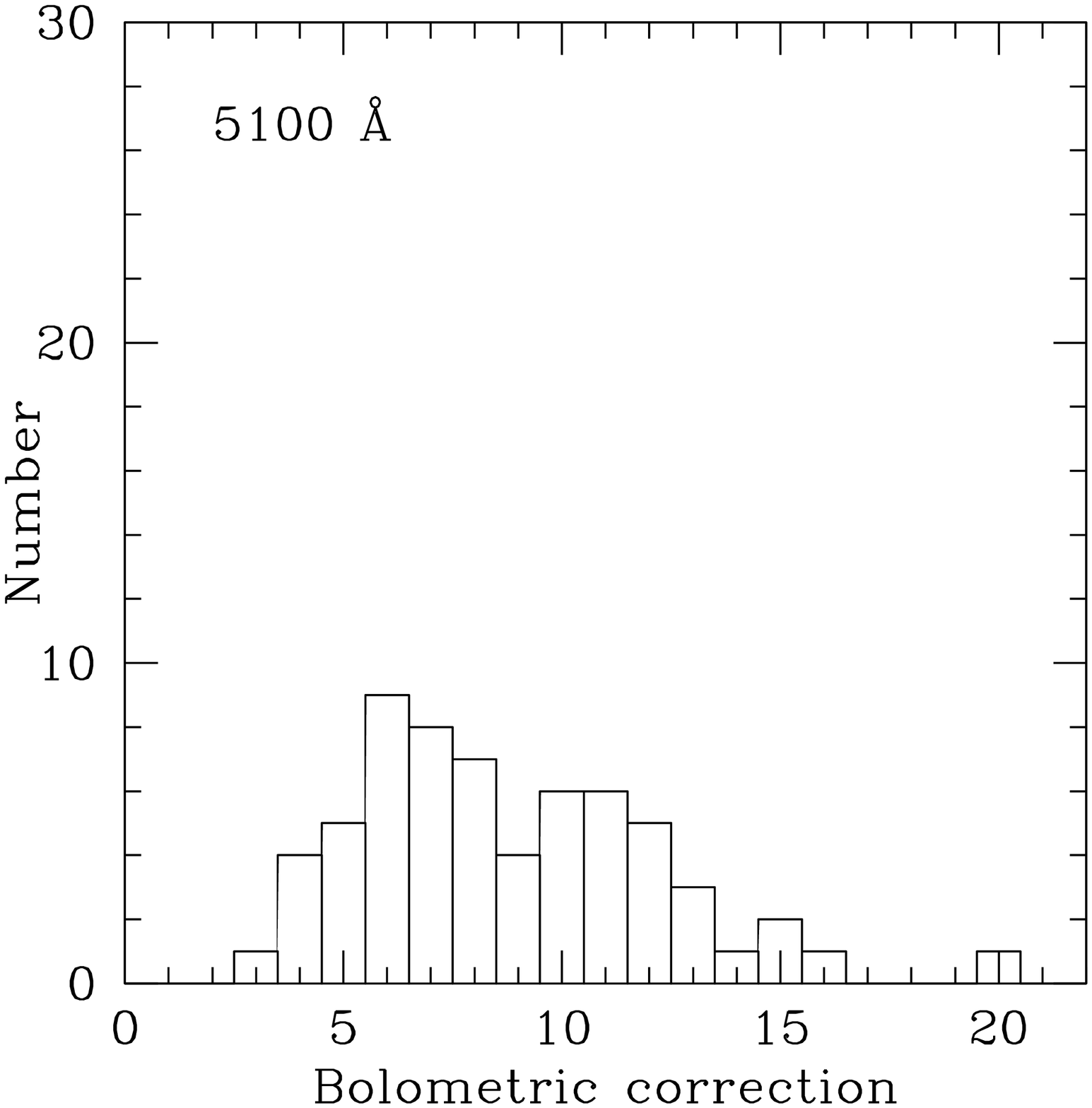}
\end{minipage}                             
\caption{Histograms showing distribution of individual linear bolometric corrections with zero intercepts at 1450, 3000, and 5100 \AA. \label{fig:hist}}
\end{figure*}

The ratios of $L_{iso}$ to $\lambda \,L_{\lambda}$ for individual objects are given in Table \ref{tab:sample}.  Fig. \ref{fig:hist} shows the distribution of values for this ratio in each wavelength window.  The `FUSE-HST' subsample is known to be UV-bright and may introduce a slight bias towards higher UV luminosities.  This would have the effect of reducing dispersion around the mean at 1450 \AA, although it is not solely responsible for this behavior.     

We found a similar behavior of bolometric correction distributions with wavelength to \citet{nb10}.  Bolometric correction distributions become tighter and shift to a lower mean with decreasing wavelength (see Fig. \ref{fig:hist}) .  This is due to variations in SED shape from object to object.  The peak of the Big Blue Bump is closer 1450 \AA\ which makes bolometric corrections there less sensitive to variations in the shape of the bump.  \citet{nb10} find mean bolometric corrections (included in Table \ref{tab:linbolcor}) of 3, 5.9, and 7.6 for 1450, 3000, and 5100 \AA\ respectively.  Note that their limits of integration are IR (3 $\mu$m) to X-ray (414 eV).  The \citet{hubeny00} models drop off in the IR and are orders of magnitude below the peak so a large difference does not result from the \citet{nb10} lower limit compared to ours.

Finally, we provide best-fitting bolometric corrections for $L_{iso}$ versus $\lambda \, L_{\lambda}$.  These and other fits in this paper are made by minimizing the chi-squared statistic using \textsc{mpfit} \citep{markwardt09} which employs the Levenberg-Marquardt least squares method.  The uncertainties in $\lambda \,L_{\lambda}$ are considered to be negligible compared to those in $L_{iso}$ because the spectra are high signal-to-noise and the monochromatic fluxes are well known.  There are many sources of uncertainty in $L_{iso}$ including, but not limited to, the interpolation, variation in SED shape, host galaxy corrections, and orientation.  While there is uncertainty in the integration process associated with our selection of a model in NIR and X-ray as discussed, we avoid more robust fitting methods and choose only to minimize the chi-squared statistic.

The resulting linear bolometric corrections are 4.2$\pm$0.1, 5.2$\pm$0.2, and 8.1$\pm$0.4 at 1450, 3000, and 5100 \AA, respectively.  The uncertainties given here are the 1-$\sigma$ errors in the fit coefficients.  These are listed in Table \ref{tab:linbolcor} and compared to other values from the literature.

\begin{table*} 
\begin{minipage}{13cm}
\caption{Bolometric Corrections for $(L_{iso}) = \zeta\, \lambda L_{\lambda}$ \label{tab:linbolcor}}
%\scalebox{0.85}{
\begin{tabular}{lcccccc}
\hline
 		 &  			      &				 &  			     & Number     & Range in            & Standard error in mean       \\
Reference & $\zeta_{1450}$ & $\zeta_{3000}$ & $\zeta_{5100}$ & of sources & log$(L_{bol})$  & for 1450/3000/5100 \AA \\
\hline
\citet{elvis94} \footnote{Recalculated from the full \citet{elvis94} radio-quiet mean SED for 1450, 3000, and 5100 \AA.} &  5.12  &  6.19  &  12.45  & 47 & 44.86-46.92 & \nodata \\
Recalculated \citet{elvis94} \footnote{Corrections are for L$_{iso}$ recalculated for the \citet{elvis94} radio-quiet mean SED from 1 $\mu$m to 8 keV.} & 3.15 & 3.82 & 7.68 & \nodata & \nodata & \nodata \\
\citet{richards06}	& \nodata &  5.62 &  10.33 & 259 & 45.06-47.43 & \nodata/0.07/0.13 \\
Recalculated \citet{richards06} \footnote{Corrections are for L$_{iso}$ recalculated for the \citet{richards06} mean SED containing all objects from 1 $\mu$m to 8 keV.} & 2.33 & 3.11 & 5.53 & \nodata & \nodata & \nodata \\
\citet{nb10} \footnote{These corrections are based on \citet{hubeny00} theoretical accretion disc models which are known to produce insufficient X-rays that drop off steeply before 0.1 keV.}	&  3.0  &  5.9  &  7.6 & 280 & 44.60-48.50 & 0.3/0.8/1.9 \\
This Work			&  4.2  &  5.2  &  8.1  & 63 & 45.13-47.30 & 0.1/0.2/0.4 \\ 
\hline
\end{tabular} 
%}
\end{minipage}
\end{table*}

There are several differences between the bolometric corrections derived here and those derived in previous studies \citep[namely][]{elvis94,richards06}.  Both integrated the full SED, \citet{richards06} specifically from $100\,\mu$m to $10\,$keV, in order to determine the bolometric luminosity.  Here we integrate only the NIR-through-X-ray region.  As a result, \citet{elvis94} and \citet{richards06} typically find larger bolometric corrections.  

Figures \ref{fig:lofit}, \ref{fig:medfit}, and \ref{fig:hifit} illustrate the difference between integration methods employed here and in the literature.  The \citet{elvis94} and \citet{richards06} bolometric corrections will systematically overestimate bolometric luminosity compared to our corrections due to their inclusion of IR emission.  The \citet{elvis94} linear relations give bolometric luminosities that are a factor of 1.22 1.19, and 1.54 greater than those derived here at 1450, 3000, and 5100 \AA, respectively.  For \citet{richards06} we find that their linear relations return bolometric luminosities a factor of 1.08 and 1.27 larger than those derived here at 3000 and 5100 \AA, respectively.

In order to make a self-consistent comparison to the \citet{elvis94} and \citet{richards06} corrections, we obtained their data and reintegrated their composite SEDs using our limits of integration. Since we only integrated the composites of \citet{elvis94} and \citet{richards06} and did not determine bolometric corrections for individual objects, we cannot compare directly to our bolometric corrections from Table~\ref{tab:linbolcor}.  Instead we derived bolometric corrections from the RQ composite spectrum from \citet{shang11} to make our comparison.  When we did this, we found the \citet{elvis94} and \citet{richards06} relations will rather underestimate bolometric luminosity.  Corrections derived from the RQ composite of \citet{elvis94} will underestimate bolometric luminosity compared to corrections derived from the RQ composite of \citet{shang11} at all luminosities by factors of 0.75, 0.76, and 0.90 at 1450, 3000, and 5100 \AA, respectively.  Corrections derived from the RQ composite of \citet{elvis94} will underestimate bolometric luminosity compared to corrections derived from the RQ composite of \citet{shang11} at all luminosities by factors of 0.55, 0.62, and 0.65 at 1450, 3000, and 5100 \AA, respectively.  The differences in these bolometric corrections reflect the differences in the SEDs from which they are derived and highlight the importance of sample selection in deriving and applying bolometric corrections.  The \citet{elvis94} and similarly the \citet{shang11} composites are from samples that are known to be UV bright and thus require larger bolometric corrections than the \citet{richards06}.  See \citet{shang11} figures 6 and 8 for a comparison of the composites.  Generally, it appears that the difference in bolometric corrections is the result of different samples as well as a different method and integration limits.

\subsection{Linear Bolometric Corrections with Non-Zero Intercepts}
We derived a second bolometric correction of the form used by \citet{nb10}:

\begin{equation}
\textnormal{log}(L_{iso}) = A+B\,\textnormal{log}(\lambda L_{\lambda}).
\end{equation}

We fit a linear relationship to the luminosity data in log-space with both $A$ and $B$ allowed to vary freely.  We found that, in general, this improved the reduced $\chi^2$ from the linear fit with a zero intercept.

At 1450 \AA, adding a nonzero intercept significantly improved the fit.  Using the regression analysis package in MiniTab$\textsuperscript{\textregistered}$ Statistical Software \footnote{www.minitab.com}, we found that the probability that the best-fitting line has a zero intercept is $4\times10^{-7}$ and the t-ratio is 4.71.  At 3000 \AA\ the fit is only marginally improved by allowing the intercept to vary from zero.  The probability that the best-fitting line has a zero intercept is 0.151 and the t-ratio is 1.45.  At 5100 \AA\ adding a nonzero intercept to the fit is an improvement.  The probability that the best-fitting line has a zero intercept is 0.004 and the t-ratio is 2.95.

In practice, this provides a moderate improvement over using other bolometric corrections.  There is no systematic over or underestimate for all luminosities between these and our bolometric corrections with a zero intercept.  Except for the 1450 \AA\ correction at the lowest luminosities, corrections from \citet{elvis94} that include IR emission will overestimate bolometric luminosity compared to our bolometric corrections with nonzero intercepts by factors of 1.08, 1.16, and 1.45, for log($\lambda \, L_{\lambda})=45.0$ at 1450, 3000, and 5100 \AA, respectively.  The bolometric corrections of \citet{richards06} that include IR emission will also overestimate bolometric luminosity compared to our bolometric corrections with nonzero intercepts by factors of 1.05 and 1.20 for log($\lambda \, L_{\lambda})=45.0$ at 3000 and 5100 \AA, respectively.

The approach that we employ is like that of \citet{nb10} whose fits are tabulated alongside ours in Table \ref{tab:nonlinbolcor}.  They choose similar limits of integration for calculating bolometric luminosity and find a nonzero intercept to be a good fit as we do here.  Note that the \citet{hubeny00} accretion disc models are theoretical and do not produce X-rays.

In comparison, the \citet{nb10} bolometric corrections show different behaviors depending on wavelength.  At 1450 \AA\ the \citet{nb10} correction systematically overestimates bolometric luminosity compared to ours, by a factor of 1.50 at log($\lambda \, L_{\lambda})=45.0$.  At 3000 \AA\ their correction underestimates ours at most monochromatic luminosities in our range, by a factor of 0.92 at log($\lambda \, L_{\lambda})=45.0$.  At 5100 \AA\ their correction underestimates ours at most monochromatic luminosities in our range, by a factor of 0.93 at log($\lambda \, L_{\lambda})=45.0$.

\begin{table*}
\begin{minipage}{20cm}
\caption{Bolometric Corrections for log$(L_{iso})=A+B$ log$(\lambda L_{\lambda})$ and First-Order Bolometric Correction \label{tab:nonlinbolcor}}
\begin{tabular}{ccc}
\hline
Wavelength & Bolometric Correction (This Work) \footnote{Uncertainties for all fits in this table are 1 $\sigma$ errors in the fit coefficients.} & \citet{nb10} Bolometric Correction \\
\hline
1450 \AA  &   log$(L_{iso})=(4.74\pm1.00)+(0.91\pm0.02)$ log$(\lambda L_{\lambda})$ &
                        log$(L_{iso})=(6.70\pm0.69)+(0.87\pm0.020)$ log$(\lambda L_{\lambda})$ \\
3000 \AA  &   log$(L_{iso})=(1.85\pm1.27)+(0.97\pm0.03)$ log$(\lambda L_{\lambda})$ &
                        log$(L_{iso})=(9.24\pm0.77)+(0.81\pm0.020)$ log$(\lambda L_{\lambda})$ \\
5100 \AA  &   log$(L_{iso})=(4.89\pm1.66)+(0.91\pm0.04)$ log$(\lambda L_{\lambda})$ &
                        log$(L_{iso})=(11.7\pm0.93)+(0.76\pm0.020)$ log$(\lambda L_{\lambda})$ \\
\hline
5100 \AA & log$(L_{iso}) = (1.02\pm0.001)\,$log$(5100\,L_{5100}) -(0.09\pm0.03) \, \alpha_{\lambda,opt}$ & \nodata \\
\hline
\end{tabular} 
\end{minipage}
\end{table*}

\begin{figure}
\begin{center}
\includegraphics[width=8.9 truecm]{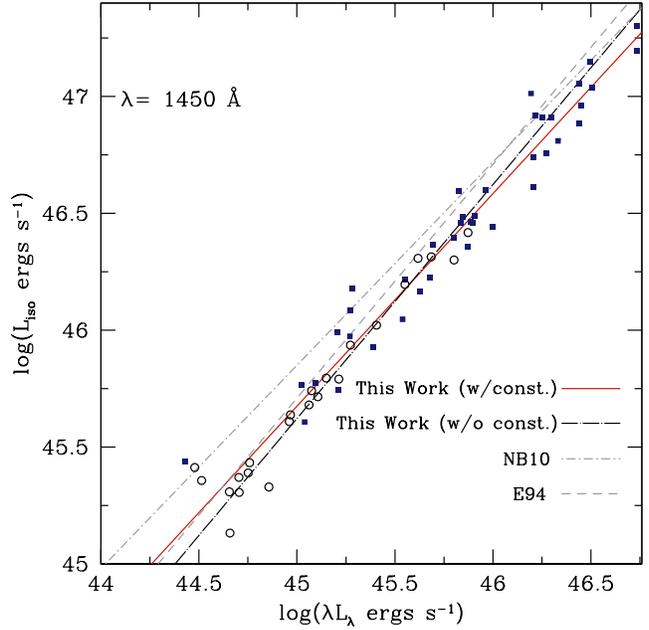}
\end{center}
\caption{Relation between $L_{iso}$ and $\lambda L_{\lambda}$ in log-log space at 1450 \AA.  The solid red line shows the $\chi^2$ fit with a nonzero intercept, the dashed-dotted black line shows our linear best-fitting bolometric correction with a zero intercept, the dashed-dotted gray line shows the linear relation with a nonzero intercept of \citet{nb10} (NB10), and the dashed gray line shows the linear bolometric correction of \citet{elvis94} (E94).  RL objects are plotted as dark blue squares and RQ objects are plotted as open black circles.}
\label{fig:lofit}
\end{figure}

\begin{figure}
\begin{center}
\includegraphics[width=8.9 truecm]{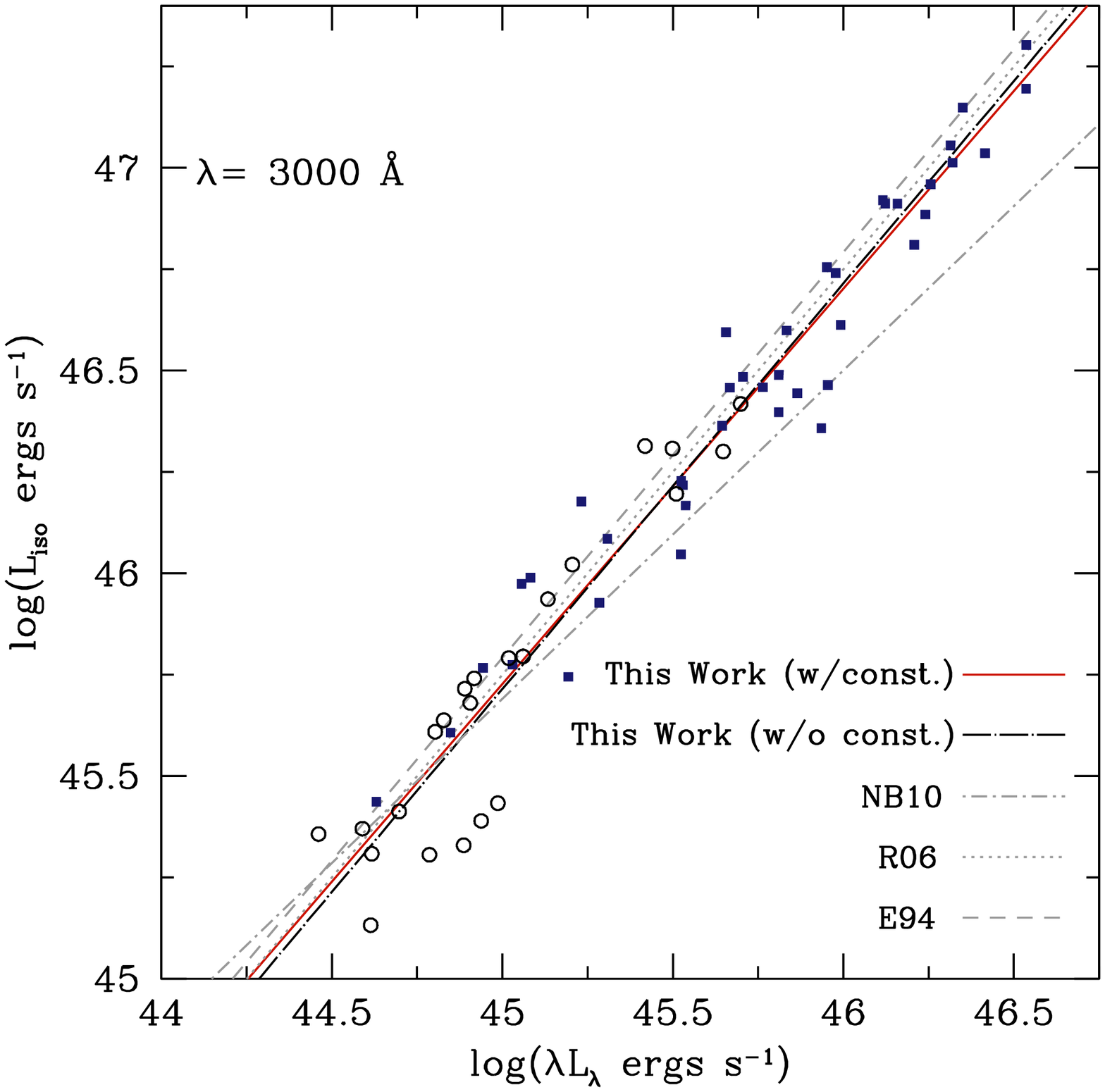}
\end{center}
\caption{Relation between $L_{iso}$ and $\lambda L_{\lambda}$ in log-log space at 3000 \AA.  The solid red line shows the $\chi^2$ fit with a nonzero intercept, the dashed-dotted black line shows our best-fitting bolometric correction with a zero intercept, the dashed-dotted gray line shows the linear relation with a nonzero intercept of \citet{nb10} (NB10), the dotted gray line shows the bolometric correction from \citet{richards06} (R06), and the dashed gray line shows the linear bolometric correction of \citet{elvis94} (E94).  RL objects are plotted as dark blue squares and RQ objects are plotted as open black circles.}
\label{fig:medfit}
\end{figure}

\begin{figure}
\begin{center}
\includegraphics[width=8.9 truecm]{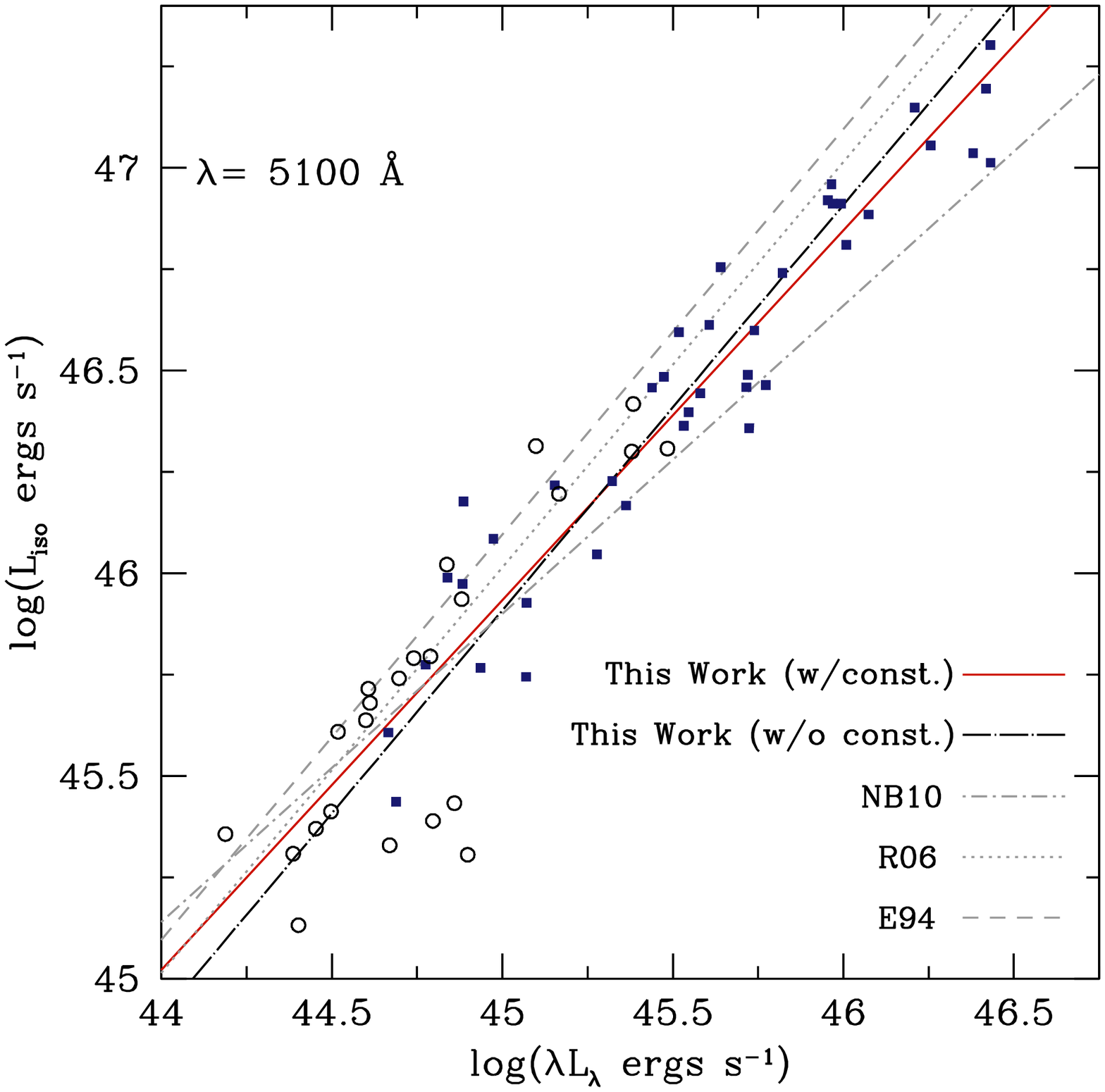}
\end{center}
\caption{Relation between $L_{iso}$ and $\lambda L_{\lambda}$ in log-log space at 5100 \AA.  The solid red line shows the $\chi^2$ fit with a nonzero intercept, the dashed-dotted black line shows our linear best-fitting bolometric correction with a zero intercept, the dashed-dotted gray line shows the linear relation with a nonzero intercept of \citet{nb10} (NB10), the dotted gray line shows the bolometric correction from \citet{richards06} (R06), and the dashed gray line shows the linear bolometric correction of \citet{elvis94} (E94).  RL objects are plotted as dark blue squares and RQ objects are plotted as open black circles.}
\label{fig:hifit}
\end{figure}

Figs. \ref{fig:lofit}, \ref{fig:medfit}, and \ref{fig:hifit} show the relationship between $L_{iso}$ and $\lambda L_{\lambda}$ in log-log space for $\lambda=1450$, $3000$, and $5100$ \AA\ respectively.  Bolometric corrections from the literature are shown for comparison.  Corrections from \citet{elvis94} and \citet{richards06} are those derived from the full SED including the IR, although those of \citet{elvis94} have been corrected to our fiducial wavelengths.

We calculated isotropic and bolometric luminosity for the 22 objects with incomplete wavelength coverage using log$(L_{iso})=(4.74\pm1.00)+(0.91\pm0.02)$ log$(1450\,L_{1450})$ from Table \ref{tab:nonlinbolcor}.  These values are presented along with other quantities of interest in Table \ref{tab:sample22}.

\begin{table*}
\begin{minipage}[2cm]{13cm}
\caption{Bolometric Luminosities for Other \citet{shang11} Objects \label{tab:sample22}}
\begin{tabular}{lccccccccccc}
\hline
Object & Redshift &  log(R*) \footnote{Radio loudness, R = f(5 GHz)/f(4215 \AA)} & $\alpha_{\lambda,opt}$ & log $(L_{iso})$ \footnote{Calculated luminosities from log$(L_{iso})=(4.74\pm1.00)+(0.91\pm0.02)$ log$(1450\,L_{1450})$.} &  log $(L_{bol})$ \footnote{Calculated luminosity adjusted for viewing angle: $L_{bol}\approx0.75\,L_{iso}$} \\
\hline
3C 110   		&  	0.7749	&      	2.68	&	-2.57		&    46.99 	& 	46.86	\\
3C 175   		&  	0.7693	&      	2.99	&  	-1.46		&    46.91 	& 	46.78	\\
3C 186   		&   	1.0630	&      	3.33	&  	-2.19		&    46.66 	& 	46.54	\\
3C 207   		&  	0.6797	&      	3.63	&  	-1.45		&    46.28 	& 	46.16	\\
3C 288.1		&     	0.9631	&      	3.42	&  	-1.87		&    46.61	&   	46.48	\\
4C 12.40		&     	0.6836	&      	3.03	& 	-0.67		&    46.17	&   	46.04	\\
4C 30.25		&      	1.0610	&      	2.92	&  	\nodata	&    46.27	&   	46.14	\\
4C 64.15		&      	1.3000	&      	3.37	&  	\nodata	&    46.73	&   	46.60	\\
B2 1351+51 	&     	1.3260	&      	2.95	&  	\nodata	&    46.66	&   	46.54	\\
B2 1555+33 	&    	0.9420	&      	2.99	&  	-2.09		&    46.21	&   	46.09	\\
B2 1611+34 	&     	1.3945	&      	3.76	&  	-1.58		&    47.06	&   	46.93	\\
MC2 0042+101&     	0.5870	&      	2.92	& 	-0.63		&    45.68	&   	45.56	\\
MC2 1146+111&  	0.8614	&      	2.55	&  	-1.01		&    46.29	&   	46.16	\\
MRK 506 		&   	0.0428	&     	0.49	&   	-1.37		&    44.67	&   	44.54	\\
PG 1103$-$006&    	0.4234	&      	2.94	&  	-1.63		&    46.30	&   	46.17	\\
PG 1259+593	&    	0.4769	&     	-1.70	&  	-1.80		&    46.83	&   	46.71	\\
PG 1534+590 	&   	0.0303	&     	0.14	&  	-0.69		&    44.43	&   	44.30	\\
PG 2214+139 	&   	0.0657	&     	-1.40	&  	-1.51		&    45.55	&   	45.42	\\
PG 2251+113 	&    	0.3253	&      	2.46	&  	-1.62		&    46.35	&  	46.23	\\
PKS 0859$-$14&      1.3320	&     	3.43	&  	-1.29		&    47.22	&   	47.10	\\
PKS 2216$-$03&    	0.8993	&      	3.23	& 	-0.47		&    46.95	&   	46.82	\\
TEX 1156+213	&   	0.3480	&      	2.38	&  	-1.51		&    46.01	&   	45.89	\\
\hline
\end{tabular}
\end{minipage}
\end{table*}

\subsection{First-Order Bolometric Corrections}
The distributions of bolometric corrections in Fig. \ref{fig:hist} and the SEDs themselves motivate an additional term in the bolometric correction.  The distribution at 1450 \AA\ is much narrower than the distribution at 5100 \AA.  The shape and wavelength of the bump vary less from object to object near the peak of the bump than on the sides so bolometric corrections at 1450 \AA, which are closer to the peak, see less variation than bolometric corrections at 5100 \AA.

In order to decrease the scatter in the 5100 \AA\ relationship, we included optical slope in a first order bolometric correction.  Optical slope helps to account for differences in SED shape by describing the relationship between the long-wavelength side and peak of the Big Blue Bump.  Optical slope is not independent of other parameters (e.g. orientation and Eddington fraction) in accretion disc models \citep{hubeny00} so the dependence of bolometric corrections on optical slope may in fact reflect these other properties to some extent.

\begin{figure}
\begin{center}
\includegraphics[width=8.9cm]{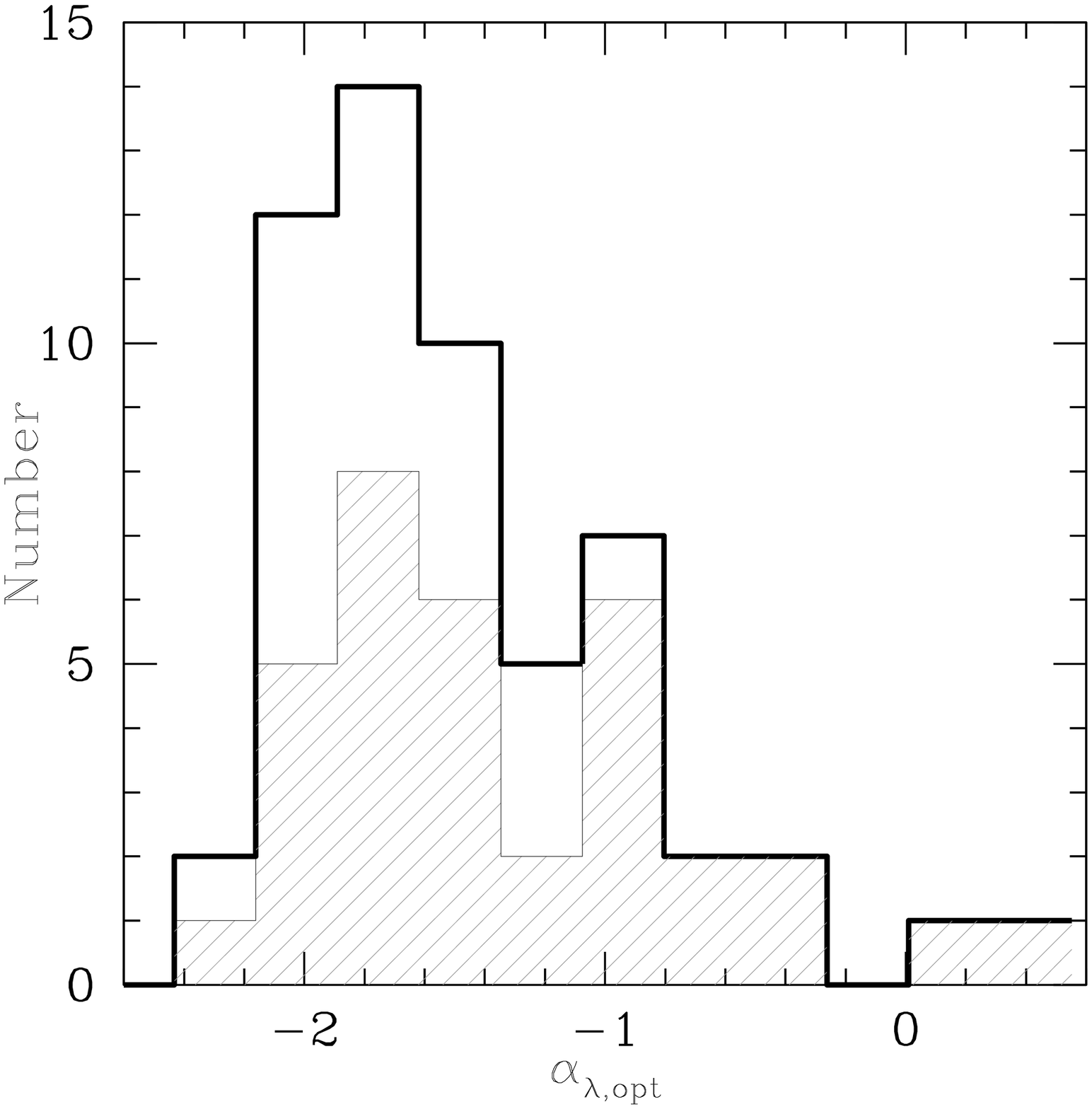}
\end{center}
\caption{Histogram of optical slopes measured in 58 objects with complete wavelength coverage.  The shaded histogram shows the distribution of optical slopes for 35 RL objects.  A K-S test gives D$_{KS}=0.12$ and a probability of 0.86 that the two distributions are from the same parent sample.}
\label{fig:alpha}
\end{figure}

We measured optical slope in most cases as a power law continuum at two points: 3981 \AA\ and 6310 \AA. In all cases, the measurement was taken under the H$\beta$ line and avoided the region contaminated by Fe\,\textsc{II} emission.  We defined $\alpha_{\lambda,opt}$ such that $F_{\lambda}\propto\lambda^{\alpha_{\lambda,opt}}$.  Because of limited wavelength coverage, optical slope could only be measured in 58 of the 63 objects with bolometric luminosities.  Fig. \ref{fig:alpha} shows the distribution of optical slopes for the full sample and only the RL portion.  The distributions of optical slope for the full sample and RL and RQ subsamples are statistically similar.  A K-S test on the full and RL sample gives D$_{KS}=0.12$ and a probability of 0.86 that the two distributions come from the same parent sample.  A K-S test on the full and RQ distributions give D$_{KS}=0.19$ and a probability of 0.54 that the two distributions come from the same parent sample.  A K-S test on the RL and RQ samples gives D$_{KS}=0.31$ and a probability of 0.10 that the two distributions come from the same parent sample.

Optical slope is provided for objects with incomplete SEDs in Table \ref{tab:sample22}.  These are not used to find the first order bolometric correction and are not included in Fig. \ref{fig:alpha}.     

Using MiniTab$\textsuperscript{\textregistered}$ Statistical Software, we performed a multiple regression analysis with $5100\,L_{5100}$ and $\alpha_{\lambda,opt}$ and found that adding optical slope is significant with a t-ratio of 2.55.  The probability that optical slope coefficient is zero is 0.014.  The bolometric correction is:

\begin{equation}
\label{eq:firstorder}
\textnormal{log}(L_{iso}) = (1.02\pm0.001)\,\textnormal{log}(5100\,L_{5100}) - \\ (0.09\pm0.03) \, \alpha_{\lambda,opt}.
\end{equation}
%this equation is too wide

The uncertainties given here are standard deviations in the coefficients.  This bolometric correction, also listed in Table~\ref{tab:nonlinbolcor}, gives bolometric luminosities larger than those derived from other 5100 \AA\ corrections given here, smaller than those of \citet{elvis94}, and similar to those of \citet{richards06}.

There are two potential concerns with including optical slope in a bolometric correction: a break in optical slope, and reddening.  \citet{vandenberk01} find that optical slope breaks near 5000 \AA\ in composites of SDSS quasars and conclude that this is partially due to contamination from the host galaxy and partially a real change in the continuum.  If this is present in our objects, it might cause us to overestimate L$_{iso}$ by including some light from the host galaxy.  \citet{vandenberk01} find an optical slope of $\alpha_{\lambda} = -1.56$ for their geometric mean composite spectrum over the wavelength range of ~1300-5000 \AA\ (shortward of the break), and a slope of $\alpha_{\lambda} = 0.45$ after the break.  We found a mean value for optical slope of $\alpha_{\lambda,opt}=-1.53\pm0.65$ so it seems likely that this issue is minor in our sample.

Bolometric corrections based on optical slope are subject to the effects of intrinsic reddening in the SEDs.  If present, reddening would cause more positive optical slopes, which would in turn underestimate true bolometric luminosity.  We have visually inspected the SEDs and do not see evidence of intrinsic reddening except perhaps in a few objects.  We do not think that reddening is significant in this sample, but it has not been systematically removed from the SEDs which should be noted when using these corrections.  

\subsection{Isotropy and Viewing Angles \label{sec:isotropy}}
Based on studies of theoretical accretion disc models, we suggest a correction to bolometric luminosity that results from our assumption of isotropy.  An accretion disc viewed face-on (i.e. at $0^{\circ}$) will be UV-brighter than one viewed edge-on.  Bolometric corrections based on face-on observations may thus overestimate bolometric luminosity.  There are intermediate angles where isotropy is a reasonable assumption (theoretically near $55^{\circ}$ according to fig. 11 of \citealt{nb10}), but quasars may not be viewed at those angles on average.  

There is evidence to suggest that, at a viewing angle of $55^{\circ}$, we would not be able to see into the central region of the AGN to view a quasar and a correction for assuming isotropy is warranted.  \citet{nenkova08b} compare dusty torus models \citep{nenkova08a} to observations of type 1 and type 2 sources and find general agreement for opening angles in the range $40^{\circ}-60^{\circ}$.  For torus opening angles in this range, a line of sight at $55^{\circ}$ would intersect the dusty torus.

If the viewing angle is known, or can at least be estimated, it would be possible to make a correction for the assumption for isotropy based on viewing angle.  \citet{barthel89} estimates an average viewing angle from his study of a sample of 3CRR \citep{laing83,spinrad85} bright radio sources with $0.5<z<1.0$ selected at 178 MHz.  At low radio frequencies, emission is dominated by optically thin synchrotron radiation from radio lobes which emit isotropically, leading Barthel to assume that the sample was not significantly biased with respect to orientation.  He used the ratio of broad-lined quasars to narrow-lined radio galaxies, presumably quasars obscured by a torus along our line of sight, to estimate an average viewing angle of quasars to be $31^{\circ}$.  This angle is determined by integrating $\int_{0^{\circ}}^{44.4^{\circ}} \theta\, \textrm{sin}(\theta)\, d\theta$, where the limits of integration are a face-on disc and the edge of the dusty torus.

If we assume that the \citet{shang11} sample is randomly selected and is preferentially viewed at the angle of $31^{\circ}$ suggested by \citet{barthel89}, bolometric luminosity will be over-estimated by about 33 per cent \citep[][fig. 11.]{nb10}.  We therefore suggest that the bolometric corrections derived here be decreased as follows:

\begin{equation}	
	L_{bol}=f\,L_{iso} \approx 0.75\, L_{iso}
\end{equation}

\noindent where $f$ is a factor describing the bias of an average viewing angle due to the anisotropic emission from a disc.  For objects to which the line of sight is known, a theoretical correction as a function of viewing angle can be found from \citet{nb10}, fig. 11 and applied to Equation \ref{eq:firstorder}.  Their predicted range for this correction to L$_{iso}$ is approximately 67 to 200 per cent for a relativistic disc including projection effects and limb darkening at viewing angles of zero to ninety degrees.

We remind the reader that our sample may not represent a random distribution of viewing angles.  The `FUSE' and `PGX' subsamples were selected to be UV-bright.  This means those subsamples may be biased to more face-on objects (i.e. smaller viewing angles).  The `RLQ' sample may be biased towards more edge-on orientations (i.e. larger viewing angles) because the blazars (which are viewed pole-on) were removed.

Since the relations of \citet{elvis94} and \citet{richards06} already overestimate bolometric luminosity compared to corrections provided here, applying the correction for viewing angle further widens the gap.  Adopting the correction for isotropy in addition to one of our bolometric corrections will give bolometric luminosities as much as a factor of two lower, on average, than those resulting from \citet{elvis94} and \citet{richards06} corrections.  As before, the difference in bolometric luminosity at 1450 and 3000 \AA\ is less than at 5100 \AA, and bolometric luminosities resulting from the corrections of \citet{richards06} are more similar to ours than those resulting from the corrections of \citet{elvis94}.

\subsection{Higher Order and Emission-Line Bolometric Corrections}
We might expect that a higher order bolometric correction may fit the data due to the relationship between optical-X-ray slope, $\alpha_{ox}$, and luminosity \citep[e.g.,][]{just07}.  We investigate the data and see no need for a higher order bolometric correction over this range at these wavelengths.  It is possible that at higher luminosities the data will depart from a linear model.

It may be possible to provide more accurate bolometric corrections based on correlations between emission-line properties and properties of the SED.  For example, \citet{kruczek11} measure C \textsc{IV} equivalent width (EW) and velocity shift (relative to the quasar rest frame) for a sample of radio-quiet SDSS quasars and determine that objects that fall on opposite ends of the EW-velocity shift continuum will require different bolometric corrections.  Specifically, bolometric corrections will be underestimated for objects with large EW and small blueshift and overestimated for objects with small EW and large blueshift.  Emission-line bolometric corrections of this type are beyond the scope of this paper and may be pursued in the future.

\subsection{Radio Class Issues\label{sec:radioclass}}
In the previous sections we have not differentiated between RL and RQ objects when deriving optical/UV bolometric corrections.  This treatment requires specific justification because RL and RQ quasars are known to have differences in their SEDs.  

The behavior of the SED at X-ray wavelengths is well known.  \citet{zamorani81} find that RL quasars are typically three times more X-ray luminous than RQ quasars with similar optical/UV luminosities.  \citet{miller11} characterize the X-ray excess of RL quasars with respect to RQ quasars and find that it can be as high as a factor of 3.4-10.7 for extremely radio-loud quasars with R$^{*}>$ 3.5.  These differences might indicate that different bolometric corrections as a function of radio class may be appropriate.  We did not assume a priori that RL and RQ quasars can be combined.

We made fits of bolometric luminosity versus monochromatic luminosity to determine whether RL and RQ require separate bolometric corrections.  Because the RL objects have higher luminosities on average, we isolated radio-loudness by fitting only the objects in the region where RL and RQ overlap in luminosity.  

\begin{figure*}
\begin{minipage}[!b]{8cm}
\centering
\includegraphics[width=8cm]{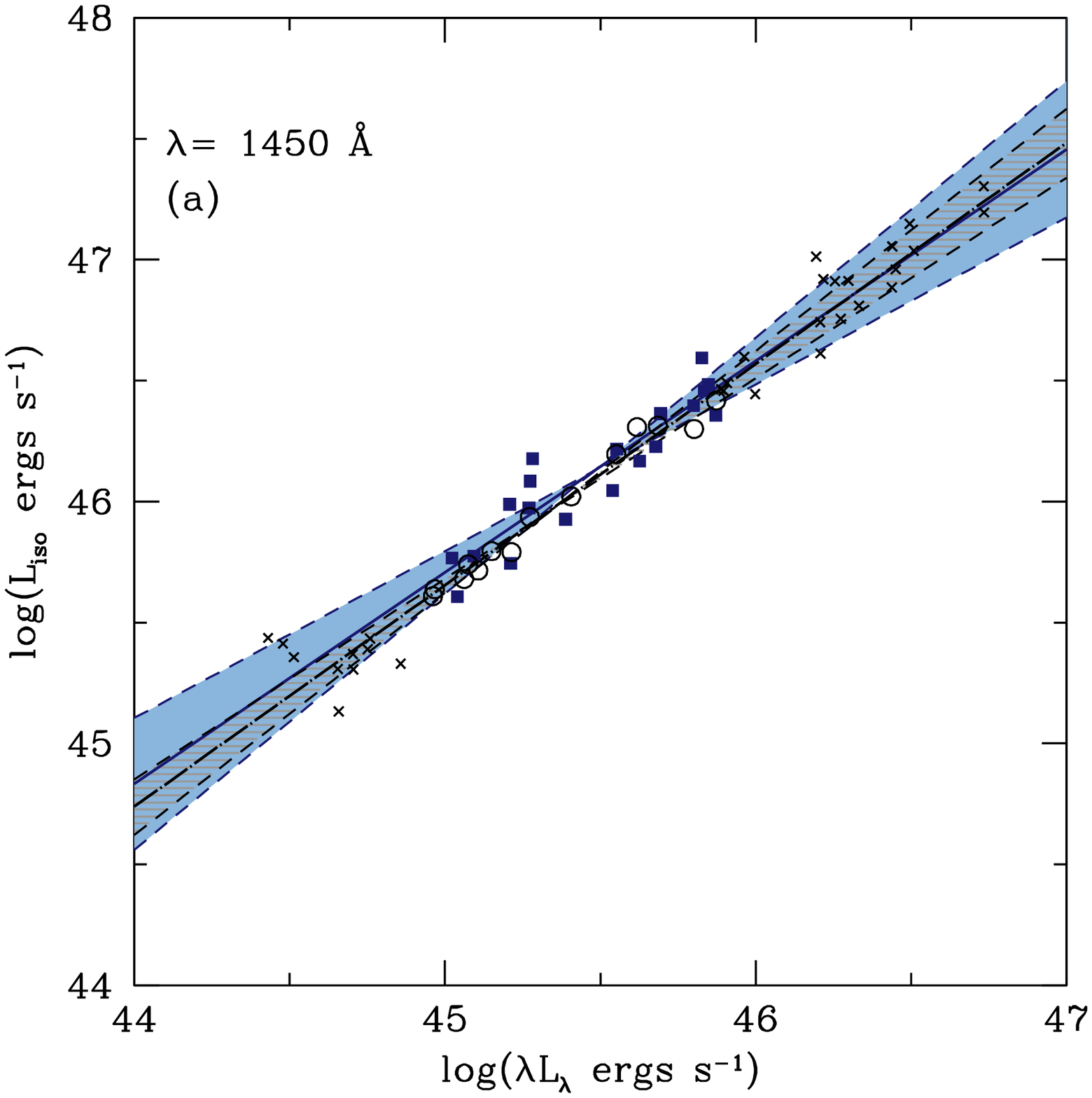}
\end{minipage}
\hspace{0.6cm}
\begin{minipage}[!b]{8cm}
\centering
\includegraphics[width=8cm]{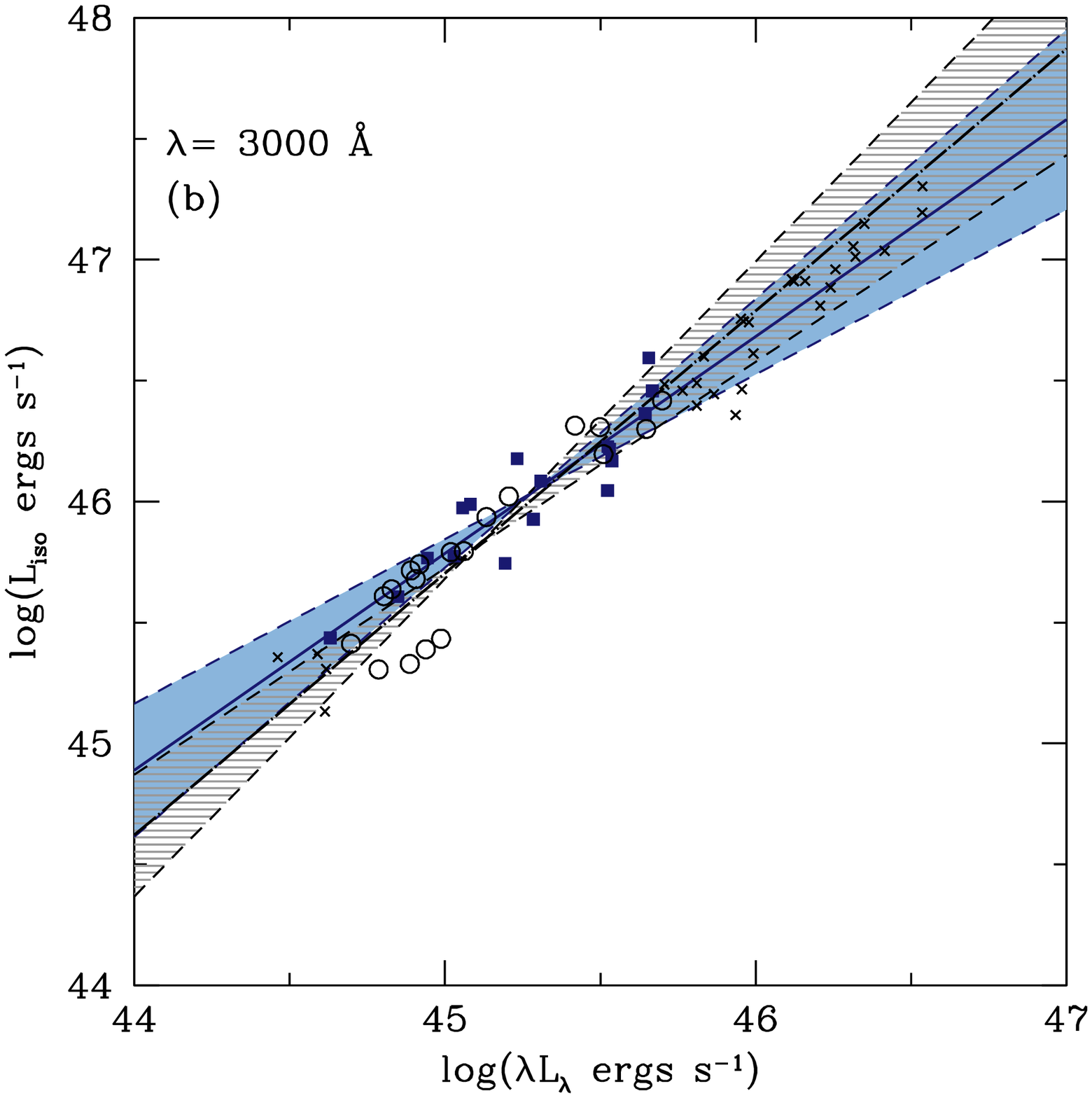}
\end{minipage}
\hspace{0.6cm}
\begin{minipage}[!b]{8cm}
\centering
\includegraphics[width=8cm]{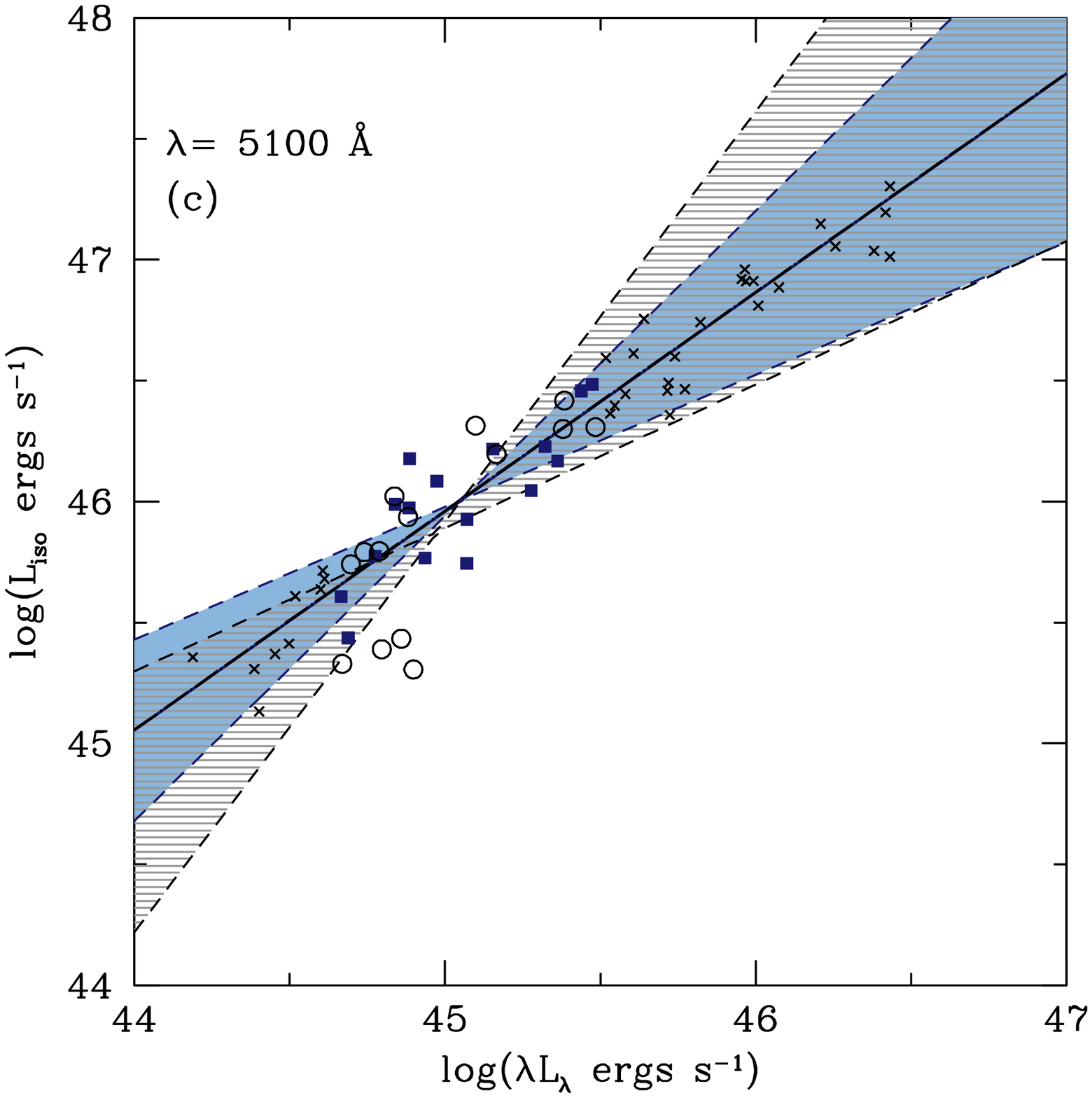}
\end{minipage}
\caption{Fits to bolometric luminosity versus monochromatic luminosity for RL and RQ quasars.  Data are used for fitting only in the region where RL and RQ quasars overlap in luminosity.  RL quasars are plotted as blue squares and the best-fitting line is solid blue.  RQ quasars are plotted as open black circles.  Black crosses are RL and RQ objects that do not overlap in luminosity.  The best-fitting line is a dashed-dotted black.  Dashed lines indicate 95\% confidence intervals.  RL quasar confidence intervals are shaded light blue, RQ quasar confidence intervals are shaded with gray cross-hatch.  Figures (a), (b), and (c) are for 1450, 3000, and 5100 \AA, respectively.  \label{fig:confidence}}
\end{figure*}

Fig. \ref{fig:confidence}, where panels (a), (b), and (c) are for 1450, 3000, and 5100 \AA respectively, shows the full sample with RL objects in the overlapping region plotted as blue squares and RQ objects in that region plotted as open black circles.  The best-fitting lines are solid blue and dashed-dotted black respectively.  The dashed lines show the 95\% confidence intervals which are shaded blue for RL and cross-hatched in gray for RQ.

Based on Fig. \ref{fig:confidence} we determined that the bolometric corrections for RL and RQ are consistent within the 95\% confidence intervals at optical/UV wavelengths.  Relaxing our fits to include all of the objects does not change this result, nor does comparing confidence intervals for fits to the RL and RQ subsamples.  This is likely because, while the X-ray behavior in RL and RQ objects is different, it is not energetically significant compared to the Big Blue Bump (which is similar in RL and RQ quasars, see the SEDs in \citealt{shang11}).  Based on this analysis, we combine RL and RQ quasars for optical bolometric corrections.

The caveat to combining RL and RQ optical/UV bolometric corrections is the unknown emission in the extreme UV and soft X-ray; RL and RQ quasars may require a different interpolation in this region.  While it is not currently possible to observe directly whether RL quasars have more ionizing flux in the EUV-soft-X-ray than RQ quasars, it may be possible to infer it from emission-line properties, though this is beyond the scope of this paper.

%X-RAY BOLOMETRIC CORRECTIONS
%%%%%%%%%%%%%%%%%%%%%%%%%%%%%%%%%%%%%%%%%%%%%%%%%%%%%%%%%%%%%%%%%%%%%%%%%%%%%%%%%
\section{X-ray Bolometric Corrections}
\label{sec:xray}
We provide several bolometric corrections to the 2-10 keV X-ray luminosity, which is calculated by extrapolating the data to 10 keV with a power law of the same slope as the highest energy X-ray data and integrating the SED between 2 and 10 keV.  As with the optical/UV corrections, we first determined whether it is appropriate to treat RL and RQ objects together.

\begin{figure*}
\begin{minipage}[!b]{8cm}
\centering
\includegraphics[width=8cm]{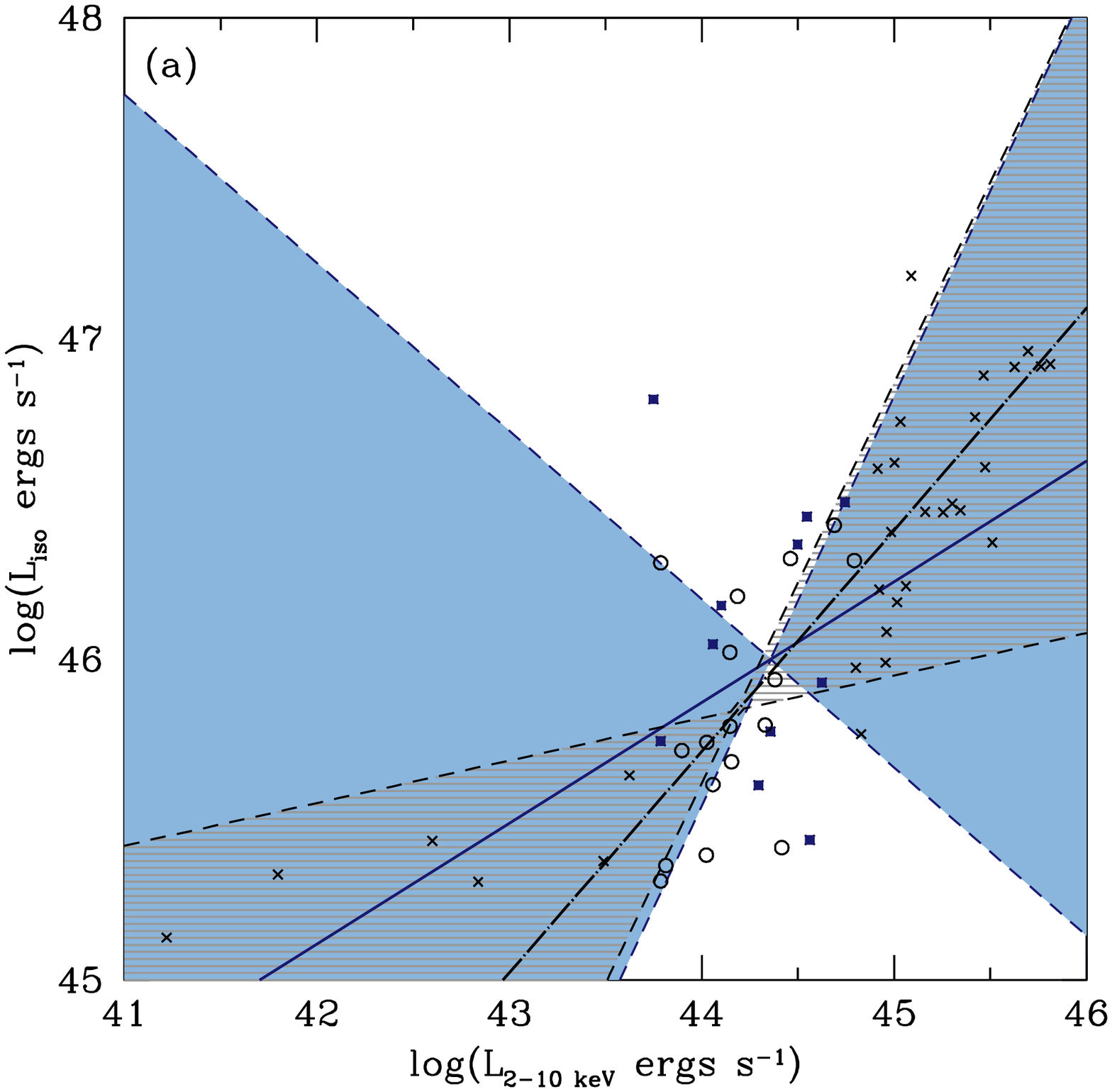}
\end{minipage}\hspace{0.6cm}
\begin{minipage}[!b]{8cm}
\centering
\includegraphics[width=8cm]{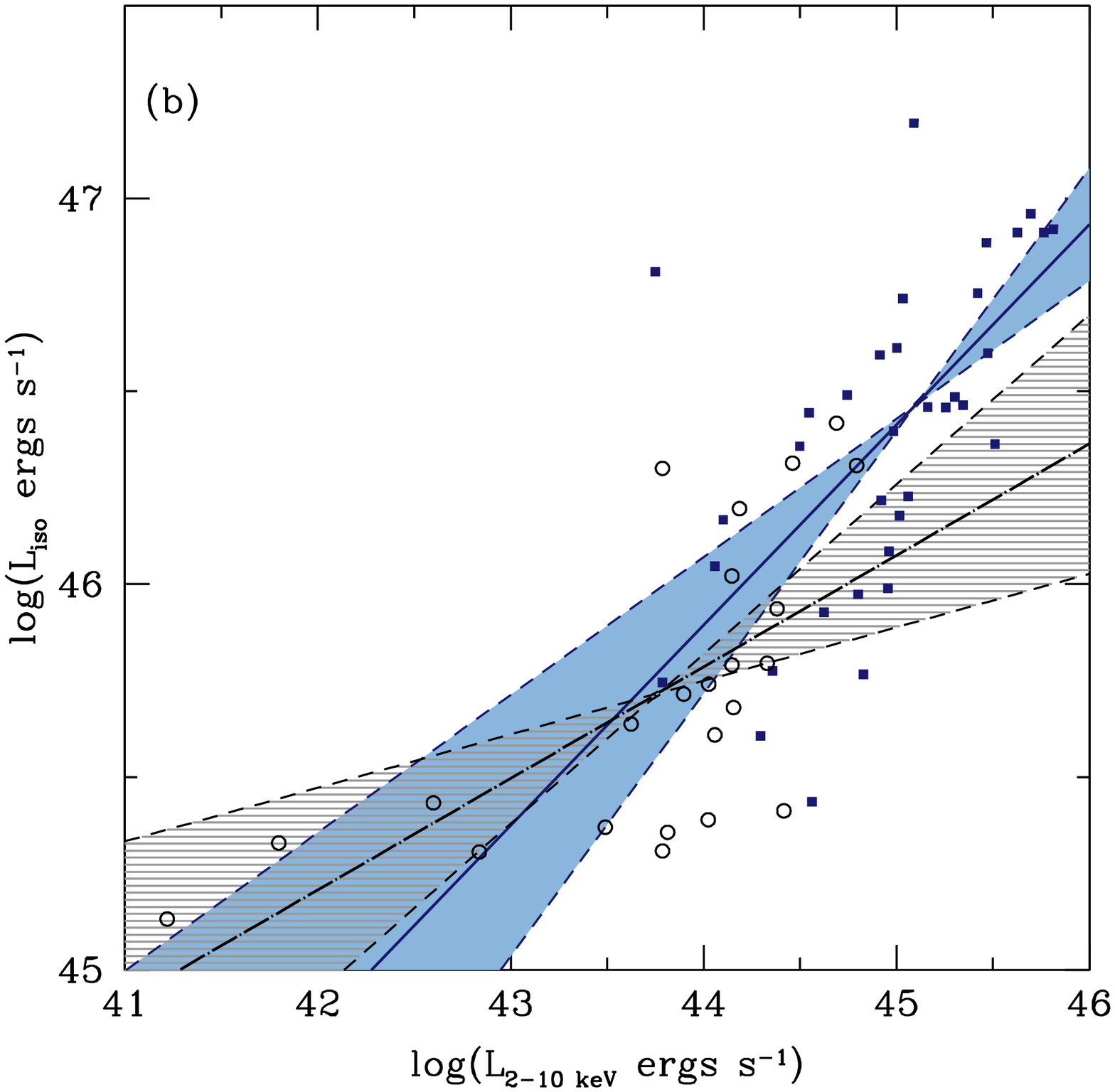}
\end{minipage}
\caption{Fits to bolometric luminosity versus X-ray luminosity for RL and RQ quasars.  In panel (a), data are used for fitting only in the region where RL and RQ quasars overlap in luminosity.  RL quasars are plotted as blue squares and the best-fitting line is solid blue.  RQ quasars are plotted as open black circles.  The best-fitting line is a dashed-dotted black.  Dashed lines indicate 95\% confidence intervals.  RL quasar confidence intervals are shaded light blue, RQ quasar confidence intervals are shaded with gray cross-hatch.  In panel (b) all of the data are used for fitting and the same conventions apply. \label{fig:Xconfidence}}
\end{figure*}

To evaluate the 2-10 keV X-ray bolometric correction for RL and RQ objects we again fit them separately and compared the fits within the 95\% confidence intervals.  Fig. \ref{fig:Xconfidence} panel (a) shows the 2-10 keV bolometric correction where we have fit the RL and RQ objects separately in the region of overlapping luminosity.  While the lines appear to differ, there are so few points in this region that the confidence intervals are very large.  Within 95\% confidence the lines are consistent.  Due to the large uncertainty in these lines, we also fit the RL and RQ data without matching luminosity range.  In Fig. \ref{fig:Xconfidence} panel (b) we see that these lines are not consistent within 95\% confidence when luminosity is not matched.  The difference in RL an RQ objects at X-ray energies has been well characterized \citep{miller11} so when making X-ray bolometric corrections we provide separate RL and RQ X-ray bolometric corrections.

%cut down (a) to y=47.5

We made bolometric corrections of the linear form with and without a zero intercept fit to the entire sample, the RL and RQ objects separately, and the RL and RQ objects separately in the region of overlapping luminosity.  The fits with the nonzero intercept always had a lower chi-squared than those where the intercept is fixed at zero and so are preferred.  More complex fits \citep[e.g.,][]{marconi04,hopkins07} are not necessary to fit these data over this luminosity range.  Table \ref{tab:xray} lists X-ray bolometric corrections.

\begin{table*} 
\caption{Bolometric Corrections for 2-10 keV Luminosity \label{tab:xray}} 
\begin{tabular}{lccc}
\hline
Sample & Linear w/ zero intercept & Linear w/ nonzero intercept \\
\hline
Full Sample 		      	& $L_{iso} = (38.00\pm06.84) \, L_{2-10 keV}$ & log$(L_{iso})=(25.14\pm01.93)+(0.47\pm0.043)$ log$(L_{2-10 keV})$ \\
Radio-Loud 		      	& $L_{iso} = (23.31\pm03.91) \, L_{2-10 keV}$ & log$(L_{iso})=(23.04\pm03.60)+(0.52\pm0.080)$ log$(L_{2-10 keV})$ \\
Radio-Quiet 		  	& $L_{iso} = (88.99\pm30.18) \, L_{2-10 keV}$ & log$(L_{iso})=(33.06\pm03.17)+(0.29\pm0.072)$ log$(L_{2-10 keV})$ \\
L-matched Radio-Loud 	& $L_{iso} = (43.86\pm13.30) \, L_{2-10 keV}$ & log$(L_{iso})=(29.30\pm18.16)+(0.38\pm0.409)$ log$(L_{2-10 keV})$ \\
L-matched Radio-Quiet 	& $L_{iso} = (45.56\pm08.38) \, L_{2-10 keV}$ & log$(L_{iso})=(15.32\pm11.69)+(0.69\pm0.264)$ log$(L_{2-10 keV})$ \\
\hline
\end{tabular} 
\end{table*}

X-ray bolometric corrections have been shown to be luminosity dependent \citep{marconi04,hopkins07,vasudevan07}.  We investigated the dependance of our X-ray bolometric corrections on luminosity via $\alpha_{ox}$.  This dependence has specifically been characterized, most recently by \citet{marchese11} for 195 X-ray selected Type 1 AGN from the XMM-\emph{Newton} serendipitous survey and also \citet{lusso10} for 545 Type 1 AGN from the XMM-COSMOS survey.  We calculated $\alpha_{ox}$ from the 2500 \AA\ luminosity using the relation from \citet{steffen06}:

\begin{equation}
\alpha_{ox}  = -0.137\cdot\textnormal{log}(L_{2500})+2.638.
\end{equation}

Our objects have luminosities within the range for which this relation was derived.  We found only a weak trend with lots of scatter of X-ray bolometric corrections with $\alpha_{ox}$, if any at all, and so do not include a luminosity-dependent bolometric correction.  The lack of a strong relationship in our data where one was found by \citet{marchese11} and \citet{lusso10} may be due to weak X-ray sources that increase the scatter in our sample.  

\begin{figure}
\begin{center}
\includegraphics[width=8.9cm]{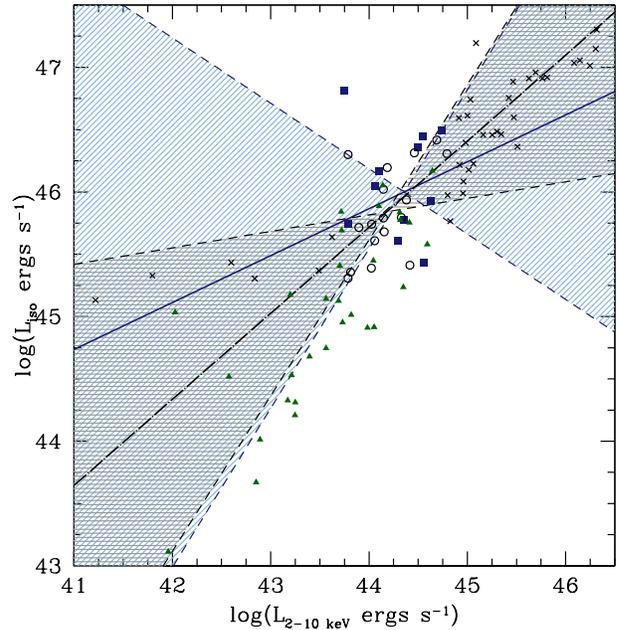}
\end{center}
\caption{Comparison between our X-ray bolometric corrections and those of \citet{vasudevan07}.  The fit to the RL data in the region of overlapping luminosity with our RQ objects, shown as blue squares, is given by the solid blue line with hashed light blue 95\% confidence intervals.  The fit to the RQ data in the region of overlapping luminosity, shown as open black circles, is given by the dashed-dotted black line with hashed gray 95\% confidence intervals.  Our high-luminosity RL objects and low-luminosity RQ objects are shown as gold crosses.  The \citet{vasudevan07} objects are shown as green triangles.  Objects of overlap between our samples have been removed from the \citet{vasudevan07} sample.  The lone RL object that sits far outside of the 95\% confidence intervals in the upper half of the plot was removed from the fit as an outlier.}
\label{fig:xcompare}
\end{figure}

\citet{vasudevan07} calculate X-ray bolometric corrections with a zero intercept for 54 AGN with model SEDs fit to optical data from \citet{baskin05} and the NASA/IPAC Extragalactic Database \footnote{The NASA/IPAC Extragalactic Database (NED) is operated by the Jet Propulsion Laboratory, California Institute of Technology, under contract with the National Aeronautics and Space Administration.} and UV/X-ray data from \emph{FUSE}.  There is some overlap between their sample and ours, but because quantities are calculated from different SEDs and limits, we found only general agreement with their X-ray bolometric corrections.  Fig. \ref{fig:xcompare} shows our data plotted with those of \citet{vasudevan07}.  Their data and ours seem to be reverse-correlated but with large scatter.  \citet{vasudevan07} point out that the X-ray bolometric corrections derived by \citet{hopkins07} and \citet{marconi04} seem to have inadequate dispersion to describe the data.  Our data support this conclusion.

\begin{figure}
\begin{center}
\includegraphics[width=8.9cm]{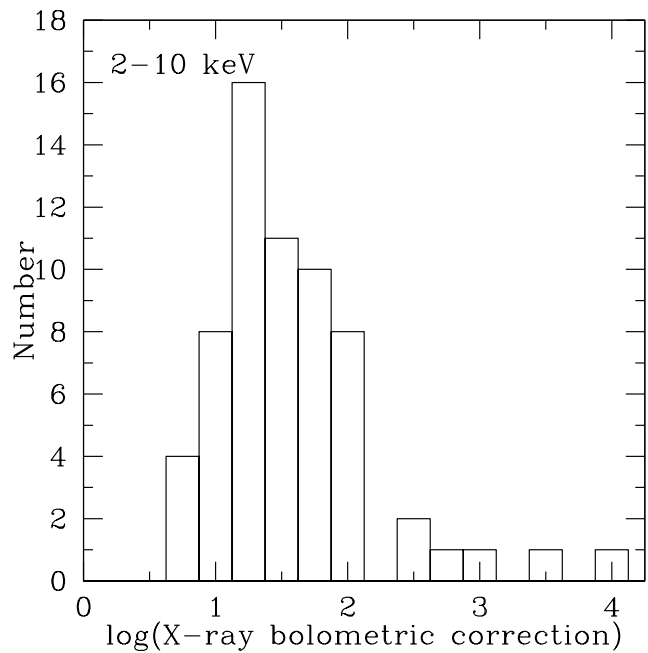}
\end{center}
\caption{Histogram of 2-10 keV X-ray bolometric corrections.  Note that the x-axis is in log in order to display the extremely wide range in X-ray bolometric corrections.}
\label{fig:xhist}
\end{figure}

We caution the reader that mean X-ray bolometric corrections are very inaccurate for individual objects due to variation in the X-ray emission from object to object.  This is in agreement with studies by \citet{marchese11} who conclude that X-ray luminosity is not a useful proxy for bolometric luminosity due to the large intrinsic scatter in X-ray bolometric corrections.  Fig. \ref{fig:xhist} shows the distribution of X-ray bolometric corrections.  The bolometric corrections are given in log in order to display the wide range of individual corrections.  

%SUMMARY
%%%%%%%%%%%%%%%%%%%%%%%%%%%%%%%%%%%%%%%%%%%%%%%%%%%%%%%%%%%%%%%%%%%%%%%%%%%%%%%%%
\section{Summary}
We provide a new generation of bolometric corrections derived from the SEDs of \citet{shang11}.  We integrated the NIR-X-ray region of each SED to estimate bolometric luminosity and derived a variety of bolometric corrections, although we determined that separate RL and RQ optical/UV corrections were not warranted by the data.  We also provided bolometric corrections to the 2-10 keV luminosity, including separate RL and RQ corrections.  

We provided optical/UV bolometric corrections with four levels of increasing sophistication and several X-ray bolometric corrections, listed below.  These are valid under the accompanying assumptions.
\begin{enumerate}
\item  Traditional, linear best-fitting bolometric corrections of the form $L_{iso}=\zeta\,\lambda L_{\lambda}$ at three wavelengths.  This assumes SED shapes do not vary from object to object and bolometric luminosity and monochromatic luminosity are linearly related.
\item  Linear, best-fitting bolometric corrections with a nonzero intercept of the form log$(L_{iso})=A+B\,\textnormal{log}(\lambda L_{\lambda})$.  
\item  A first-order bolometric correction.  The bolometric correction at 5100 \AA\ can be improved by adding a correction based on the optical continuum slope.
\item  Bolometric corrections assuming isotropy should probably be corrected for an average viewing angle bias.
\item  Bolometric corrections to the 2-10 keV luminosity for the full sample, RL and RQ subsamples, and RL and RQ subsamples in the region of overlapping luminosity.
\end{enumerate} 

Mean bolometric corrections derived here are smaller on average than those derived previously \citep[e.g.][]{elvis94,richards06} due do a difference in methodology and data.  We found good agreement with the theoretical work of \citet{nb10} as the distribution of bolometric corrections tightens and shifts to a lower mean with decreasing wavelength.

For objects with properties that fall in the region of parameter space covered by this sample (log(L$_{bol}$)=45.1-47.3), we recommend the following bolometric corrections:  

\begin{itemize}
\item {\bf Recommended Correction at 1450 \AA}
	\begin{equation}
	\label{eqn:const14}
	\textnormal{log}(L_{iso})=(4.74\pm1.00)+(0.91\pm0.02)\,\textnormal{log}(1450L_{1450})
	\end{equation}
\item {\bf Recommended Correction at 3000 \AA}
	\begin{equation}
	\label{eqn:const30}
	\textnormal{log}(L_{iso})=(1.85\pm1.27)+(0.98\pm0.03)\,\textnormal{log}(3000L_{3000})
	\end{equation}
\item {\bf Recommended Corrections at 5100 \AA}
	\begin{equation}
	\label{eqn:const51}
	\textnormal{log}(L_{iso})=(4.89\pm1.66)+(0.91\pm0.04)\,\textnormal{log}(5100L_{5100})
	\end{equation}
	\begin{equation}
	\label{eqn:opt}
	\textnormal{log}(L_{iso}) = (1.02\pm0.001)\,\textnormal{log}(5100\,L_{5100}) - (0.09\pm0.03) \, \alpha_{\lambda,opt}
	\end{equation}
\item {\bf Correction for Average Viewing Angle} \\
	This correction should be applied to bolometric luminosities resulting from any of our other optical/UV bolometric corrections.
	\begin{equation}
	\label{eqn:iso}
	L_{bol}=f\,L_{iso} \approx 0.75\, L_{iso}
	\end{equation}
\item {\bf X-ray Bolometric Corrections}
	\begin{equation}
	\textnormal{log}(L_{iso,RL})=(23.04\pm03.60)+(0.52\pm0.080)\,\textnormal{log}(L_{2-10 keV})
	\end{equation}
	\begin{equation}
	\textnormal{log}(L_{iso,RQ})=(33.06\pm03.17)+(0.29\pm0.072)\,\textnormal{log}(L_{2-10 keV})
	\end{equation}
\end{itemize}

Bolometric corrections with nonzero intercepts are marginally preferred to those without.  At 5100 \AA, the bolometric correction with optical slope is preferable to the one with the nonzero intercept.  All bolometric luminosities derived here assume isotropy and we recommend they be corrected for an average viewing angle bias.  X-ray bolometric corrections are uncertain and optical/UV corrections are preferred when possible.

%ACKNOWLEDGEMENTS
%%%%%%%%%%%%%%%%%%%%%%%%%%%%%%%%%%%%%%%%%%%%%%%%%%%%%%%%%%%%%%%%%%%%%%%%%%%%%%%%%
\section*{Acknowledgments}
The authors would like to acknowledge Beverly Wills, Derek Wills, Sabrina Cales, Daniel Dale, Richard Green, Rodrigo Nemmen, Sarah Gallagher, Rajib Ganguly, Dean Hines, Benjamin Kelly, Gerard Kriss, Jun Li, Baitian Tang, and Yanxia Xie for their considerable efforts in compiling the SED sample.

Z. S. acknowledges support by the National Natural Science Foundation of China through Grant No. 10773006 and Chinese 973 Program 2007CB815405.  We are also grateful for support by NASA through grant HST-GO-10717.01-A, Spitzer-GO-20084, and Grant No. NNG05GD03G.

\bsp
\label{lastpage}
\end{document}